\documentclass[11pt]{article} 
\usepackage[top=2cm, bottom=1.5cm, left=2cm, right=2cm]{geometry}
\usepackage{graphicx}
\usepackage[T1]{fontenc}
\usepackage{times}
\usepackage{multirow}
\usepackage{authblk}
\usepackage[utf8]{inputenc}
\usepackage{tikz}
\usepackage{tikz-3dplot}
\usepackage{hyperref}

\usepackage[caption=false]{subfig}

\usepackage{color}

\begin{document}
 \title{Graphlet characteristics in directed networks} 

\author[1]{Igor Trpevski}
\author[1]{Tamara Dimitrova}
\author[1]{Tommy Boshkovski} 
\author[1,2,3,*]{Ljupco Kocarev}
\affil[1]{Macedonian Academy of Sciences and Arts, Skopje, Republic of Macedonia}
\affil[2]{Faculty of Computer Science and Engineering, UKIM, Skopje, Republic of Macedonia} 
\affil[3]{BioCircuits Institute, UC San Diego, La Jolla, CA 92093-0402, USA}
\maketitle

A number of network structural characteristics have recently been the subject of particularly intense research, including degree distributions \cite{barabasi-1999}, community structure \cite{girvan-2002,loreto-2004}, and various measures of vertex centrality \cite{watts-1998,chung-2002}, to mention only a few. Vertices may have attributes associated with them; for example, properties of proteins in protein-protein interaction networks \cite{kim-2014}, users' social network profiles \cite{wellman-2001}, or authors' publication histories in co-authorship networks \cite{newman-2004}. In a network, two vertices might be considered similar if they have similar attributes (features, properties), or they can be considered similar based solely on the network structure. Similarity of this type is called structural similarity, to distinguish it from properties similarity, social similarity, textual similarity, functional similarity or other similarity types found in networks.  Here we focus on the similarity problem by computing (1) for each vertex a vector of structural features, called \textit{signature vector}, based on the number of graphlets associated with the vertex, and (2) for the network its \textit{graphlet correlation matrix}, measuring graphlets dependencies and hence revealing unknown organizational principles of the network. Real-world networks are directed and often have few types of vertices with well defined structural (topological) characteristics.  We found that real-world networks generally have very different structural characteristics resulting in different graphlet correlation matrices. In particular, the graphlet correlation matrix of the brain effective network is computed for 40 healthy subjects and common (present in more than 70\% subjects) dependencies are raveled. Thus, negative correlations are found for 2-node graphlets and 3-node graphlets that are wedges and positive correlations are found only for 3-node graphlets that are triangles. Graphlets characteristics in directed networks could further significantly increase our understanding of real-world networks.

Quantifying similarity between vertices in a network is an old problem. Researches have proposed comparable vertex similarity measures predicating on the idea that vertices are similar if their neighbors are similar. In social network analysis, Katz centrality \cite{katz-1953} of a vertex equals to the sum of that vertex's similarities to every other vertex and is computed by taking into account the total number of walks between a pair of vertices. In computer science, Jeh and Widom \cite{jeh-2002} have proposed a method that they call ``SimRank,'' implementing a recursive definition of object similarity: two objects are similar if they relate to similar objects. SimRank similarity of two nodes $i$ and $j$ has an elegant random walk interpretation as the probability that two independent simultaneous random walkers, beginning at $i$ and $j$, will eventually meet at some node. In physics, another similar concept which led to a self-consistent matrix formulation of similarity that can be evaluated iteratively using only the knowledge of the graph adjacency matrix is proposed by Newman and co-workers \cite{leicht-2006}.

Here we measure structural similarity between vertices on directed network by considering \textit{directed graphlets} (induced sub-graphs) associated with vertices, and, subsequently probe the structure of graphs by computing  correlations of the frequencies of different graphlets present in the graph.  Let $G=(V,E)$ be a simple directed (un-weighted) graph and let $\mathbf{A} = [a_{ij}]$ be its $n\times n$ adjacency matrix, where $n$ is the number of vertices, such that $a_{ii}=0$ (self-arcs are excluded). For an element of the edge set $E$ we write $(i, j)$ to denote $i \rightarrow j$.  A graphlet $G_k = (V_k,E_k)$ is an induced sub-graph that consists of a subset of $k$ vertices of the graph $G = (V,E)$ (i.e., $V_k \subset V$ ) together with all the edges whose endpoints are both in this subset (i.e., $E_k = \{ e \in E: e = (u, v) \mbox{ and } u, v \in V_k\}$). Therefore, graphlets are small induced sub-graphs of a large network that appear at any frequency and hence are independent of a null model. Here, we consider only weakly connected graphlets. 

Graphlets have found numerous applications as building blocks of network analysis in various disciplines ranging from social science \cite{holland-1976,faust-2010} to biology \cite{milo-2002,sporns-2004,przulj-1,przulj-2}. In social science, graphlet analysis (known as sub-graph census) is widely adopted in sociometric studies \cite{holland-1976}. Much of the work in this vein focused on analyzing triadic tendencies as important structural features of social networks (e.g., transitivity or triadic closure) as well as analyzing triadic configurations as the basis for various social network theories (e.g., social balance, strength of weak ties, stability of ties, or trust \cite{granovetter-1983}). Directed graphs are treated as having two different types of edges: directed and reciprocal. The research on reciprocal edges originates with the triad census work of Holland and Leinhardt \cite{holland-1976}. A reciprocal edge is technically a pair of directed edges, $\{(i, j), (j, i)\}$, that we treat as a single reciprocal edge. We define the set of in-neighbors $S^{-}_i = \{j:  (j,i) \in E \mbox{ and } (i,j) \notin E \} $, the set of out-neighbors  $S^{+}_i = \{j: (i,j) \in E \mbox{ and } (j,i) \notin E \} $, and the set of reciprocal-neighbors $S^\circ_i = \{j: (i,j) \in E \mbox{ and } (j,i) \in E \} $.

For a node $i$ of network $G$, the automorphism orbit of $i$ is the set of nodes of $G$ that can be mapped to $i$ by an automorphism, an isomorphism of a network with itself; i.e., a bijection of nodes that preserves node adjacency. Automorphism orbits have been used to generalize the graph degree distribution \cite{przulj-1,przulj-2}.  Given a family of graphlets of size $k$ nodes, $2 \leq k \leq 4$, one can count the number of times a node \textit{is touched} by an orbit.  In directed graphs there are 1695 orbits for up to 4-node graphlets, see SI. The larger number of such orbits makes a complete enumeration unattractive, a problem which continues to worsen for larger graphlets. Therefore, here we restrict ourselves to working with 3-node graphlets, reducing the number of orbits to 48. In addition, we further reduce it to 16 orbits by considering only isomorphism triangle and wedge classes \textit{starting} at vertex $i$ (see SI).

\begin{figure}[t]
\centering
\includegraphics{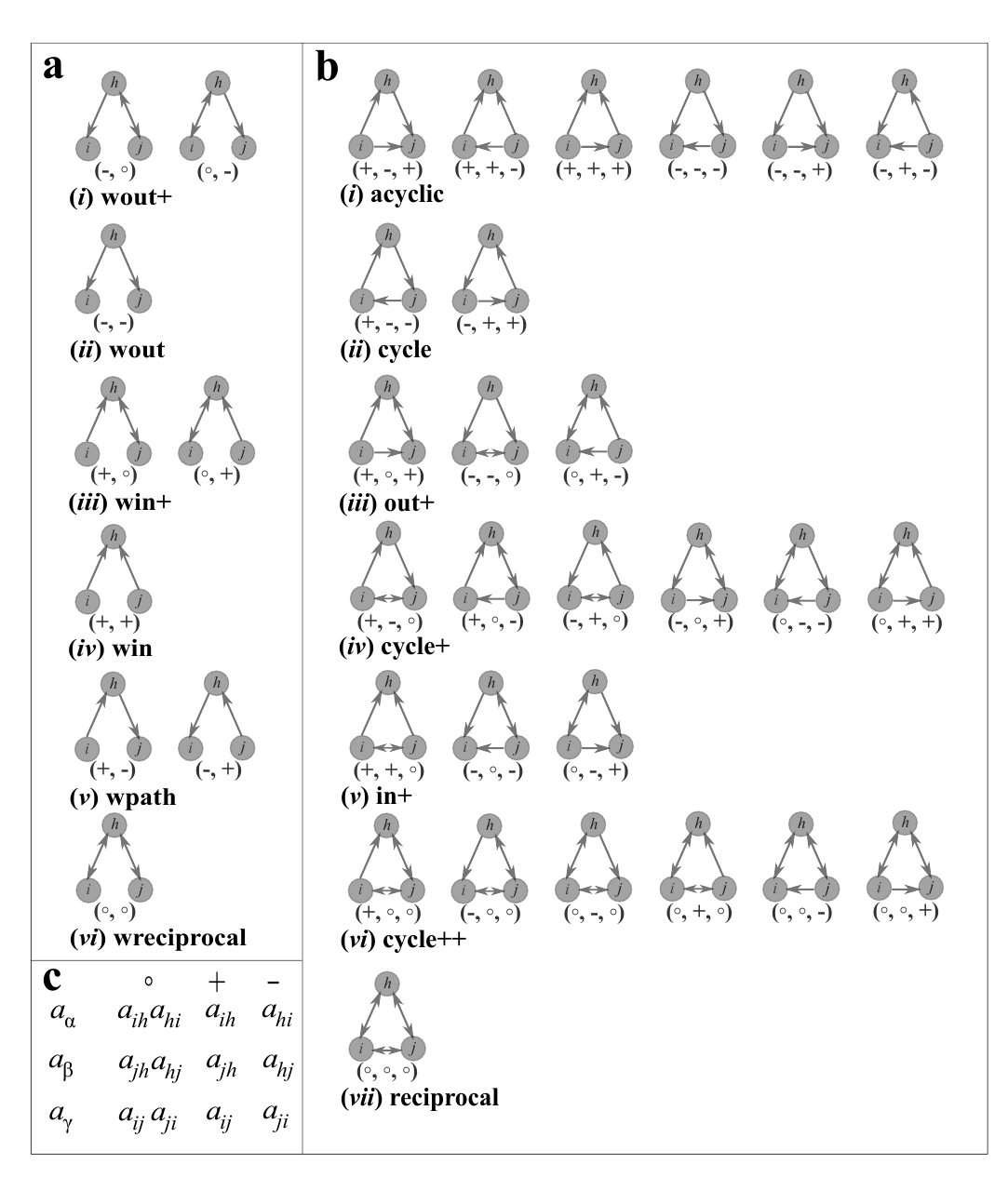}
\caption{Isomorphic classes of wedges (a) and triangles (b) in directed graphs.  In each wedge/triangle one vertex is labeled $i$ (wedge/triangle starts at $i$). Assuming that directed (and reciprocal) edges are considered with respect to particular vertex in the wedge or the triangle (see the main text), each wedge and triangle can be labeled as $(\alpha, \beta)$ and $(\alpha, \beta, \gamma)$, respectively, where $\alpha, \beta, \gamma \in \{+,-,\circ\}$. Hence, there are 9 wedges and 27 triangles starting at $i$, which are clustered in 6 wedge isomorphic classes (a) and 7 triangle isomorphic classes (b). (c) Entries of adjacency matrix for out-, in-, and reciprocal-edges.}
\label{fig1}
\end{figure}

Triangle is a set of three nodes, each of which is connected to the other two. Let $T=(V_t, E_t)$ be a triangle with nodes $i, h, j$ and edges $e_1, e_2, e_3$  between nodes $i$ and $h$, $h$ and $j$, and $i$ and $j$ respectively. We equip the edges with an arbitrary orientation, as this is necessary for the further analysis. To be specific, we assume that directed or reciprocal character of the edges between $i$ and $h$ and $i$ and $j$ are considered with respect of the node $i$ and of the edge between $h$ and $j$ with respect of the node $j$. Thus, for example, we write $(+,-, \circ)$ triangle to denote the triangle $T=(V_t, E_t)$ with $V_t = \{i,h,j\}$ and $E_t = \{  i  \rightarrow h,  h \rightarrow j, i \leftrightarrow j \}$. More generally we write $(\alpha, \beta, \gamma)$ triangle to denote one of the triangle types, where   $\alpha, \beta, \gamma = \{\circ, +, -\}$. In a similar fashion we denote $(\alpha, \beta)$ wedges for the wedges in directed graphs. Since directed graphs have two types of edges: directed and reciprocal, there exist 9 wedge types, which are arranged in 6 isomorphism wedge classes (Figure 1a), and 27 triangle types, which can be grouped in 7 isomorphism triangle classes (Figure 1b). Finally, the 2-node graphlets orbits originating at a node $i$ are equivalent to the out-links, in-links and reciprocal links of node $i$.

The number of  $(\alpha, \beta, \gamma)$ triangles \textit{starting} at $i$ is given by:
\begin{eqnarray*} \label{tri}
T_i(\alpha, \beta, \gamma) & =& \sum_{j \in S_i^{\gamma}} | S_i^{\alpha} \cap S_j^{\beta} | = \sum_{h,j; h\neq j \neq i} a_\alpha a_\beta a_\gamma
\end{eqnarray*} 
where the entries $a_\alpha, a_\beta, a_\gamma$ of the graph adjacency matrix are provided in Figure 1c. Indeed, $| S_i^\alpha \cap  S_j^\beta| $ is the number of the common $\alpha$-neighbors of $i$ and $\beta$-neighbors of $j$. Summarizing $| S_i^\alpha \cap  S_j^\beta| $ for all $j$ that are $\gamma$-neighbors of $i$, one computes the number of  $(\alpha, \beta, \gamma)$ triangles. The number of of $(\alpha, \beta)$  wedges starting at vertex $i$ is $L_i^{(\alpha, \beta)}  =  \sum_j \left| S_i^\alpha \cap S_j^\beta \right|$; the number of $(\alpha, \beta)$ wedges that are graphles is given by ${W}_i^{(\alpha, \beta)}  =  L_i^{(\alpha, \beta)} - \sum_\gamma T_i^{(\alpha, \beta, \gamma)}$.   Therefore 39 quantities $d_i^\alpha = | S_i^\alpha | $, $W_i^{\alpha, \beta}$, and $T_i^{\alpha, \beta, \gamma}$ are associated with a vertex $i$. This number is further reduced to 16 by considering only isomorphic wedge and triangle classes (SI). We define \textit{signature/feature} vector of a vertex $i$ as  $F_i = \left[  d_i, W_i, T_i \right]^T$ where: 
\begin{eqnarray*}
d_i & = & \left[ d_i^+, d_i^-, d_i^\circ \right]^T \\
W_i & = & \left[ W_i^{(path)}, W_i^{(in)}, W_i^{(out)}, W_i^{(in+)}, W_i^{(out+)}, W_i^{(rec)} \right]^T \\
T_i & = & \left[ T_i^{(acyclic)}, T_i^{(cycles)}, T_i^{(out+)}, T_i^{(cycles+)}, T_i^{(in+)}, T_i^{(cycles++)}, T_i^{(rec)} \right]^T 
\end{eqnarray*} 
$d_i^+, d_i^-, d_i^\circ$ are the numbers of 2-node graphlets, or equivalently, the out-degree, in-degree and reciprocal-edge degree of the vertex, while $W_i$ and $T_i$, called \textit{wedge-degree} and \textit{triangle-degree}, respectively, are the number of 3-node graphlets \textit{starting at} $i$. Signature vectors describe only local connectivity of directed edges.  Indeed, by flipping the direction of edges between all connected pairs of nodes with probability one-third (for out-,in-, and reciprocal-edges),  which keeps the undirected connectivity and thus the edge density unchanged, the information about the direction of edges will be destroyed and normalized signature vectors will converge to  $[\frac{1}{3}, \frac{1}{3}, \frac{1}{3}, \frac{1}{6} \ldots \frac{1}{6}, \frac{1}{7}, \ldots \frac{1}{7}]^T$ (see SI).     

\begin{figure*}[htp]
\centering
\includegraphics[width=\textwidth]{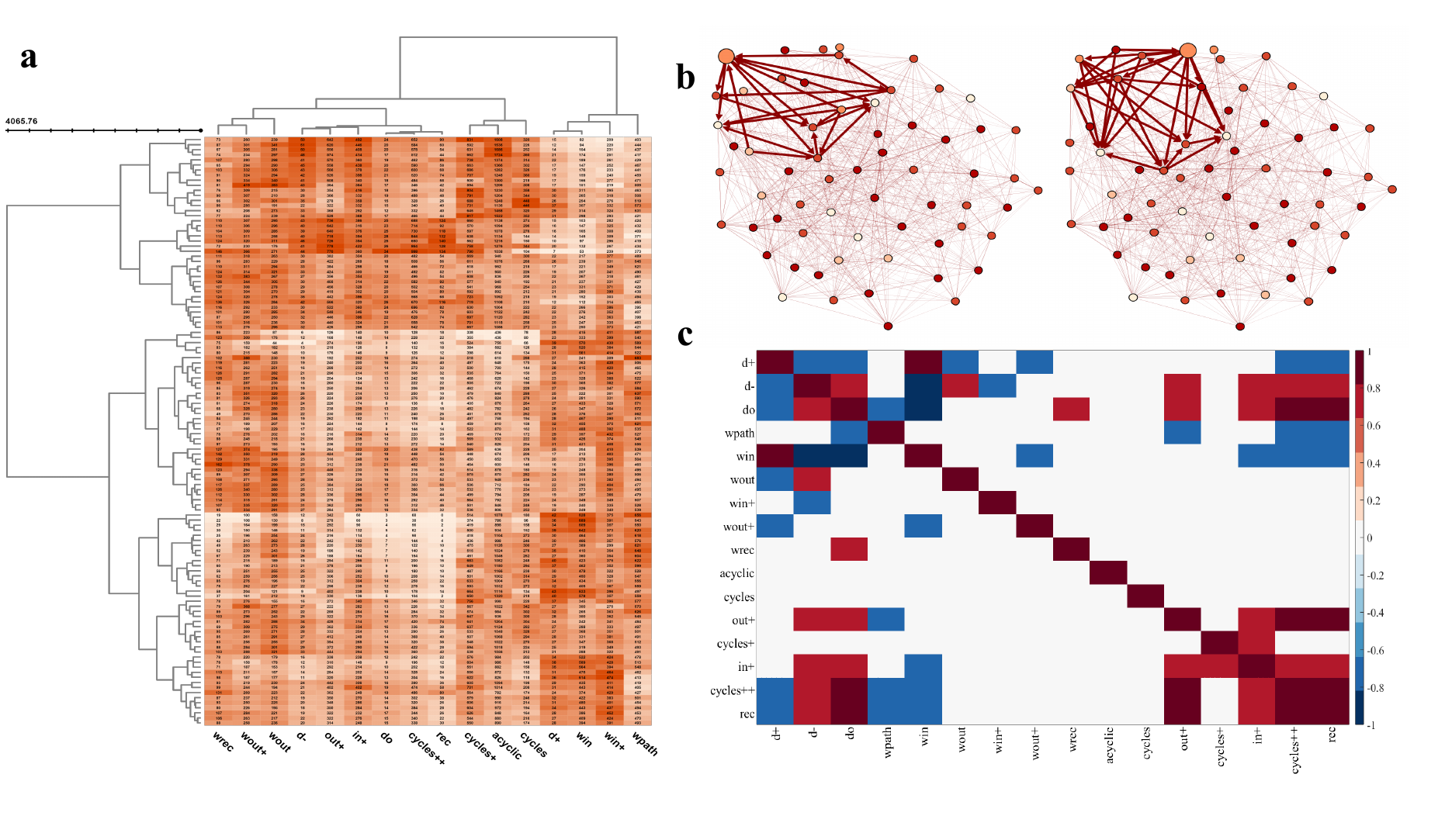}%
\caption{Graphlets in effective brain network. The dataset represents effective connectivity of the brain network, describing a network of directional effects of one neural region over another. (a) 116 regions (vertices) are considered in total (shown on the vertical axis); a directed edge represents causality of one region (vertex) over another. Each region is represented by 16-dimensional feature vector (shown on horizontal axis). Ward agglomerative hierarchical clustering procedure results in a heat-map representing clustered regions. Three clusters with similar local structure are easily indentified. (b) Two regions from the same cluster are shown, both having similar local structure; only for better visualization, sub-graph with 64 nodes is shown on which the local structures around two (similar) vertices are shown. (c) Graphlet correlation matrix of a particular brain network is shown, only those entries for which correlations (or anti-correlations) are significant ($>0.7$ and $< -0.7$) are colored.}
\end{figure*}

For each vertex in a network, we construct its signature vector consisting of 16 coordinates corresponding to the number of 16 graphlets starting at the vertex. Then we construct an $N\times16$ matrix whose rows are the signature vectors for each vertex. For a given network $G$, we compute Pearson correlation coefficients between all pairs of columns of the above described matrix and present them in a $16 \times 16$ symmetric matrix that is termed \cite{przulj-2}  \textit{graphlet correlation matrix} of network $G$. In this way, the network topology and its local direction patters, regardless network size (the number of vertices) and network volume (the number of edges), are summarized into a $16 \times 16$ matrix.   

Different networks generally have very different graphlet dependencies and hence reveal unknown organizational principles of real-world networks. We study effective brain networks of 40 healthy subjects (see supplementary information). Brain connectivity refers to a pattern of anatomical links (``anatomical connectivity''), of statistical dependencies (``functional connectivity'') or of causal interactions (``effective connectivity'') between distinct units -- region of interests (ROI) -- within a nervous system \cite{olaf}. Here, effective brain connectivity is studied using  Granger (G-) causality. Resting-State fMRI Data Analysis Toolkit (REST) is used and multivariate-coefficients ROI-wise G - causality analysis (GCA) is adopted to generate effective networks. REST-GCA supports signed-path coefficients multivariate GCA and integrates a batch mode coefficients computation for ROI-wise GCA \cite{song-2011}. For each network the number of vertices (regions) is equal to 116, while the total number of edges in the networks (after pruning) is $3466 \pm  43$. We construct a matrix such that the number of rows in the matrix is equal to the number of vertices in the network and the number of columns equals the dimension of the signature vector (the matrix has 16 columns), see Figure 2a.  The signature vector represents the local structure around the vertex and can be employed to understand organizational rules within a network and to reveal structural differences between networks. We also compute correlation coefficients between all pairs of columns of this matrix and present them in a $16 \times 16$ symmetric matrix, Figure 2c.

\begin{figure}[ht]
\begin{center}
\includegraphics[width=\textwidth]{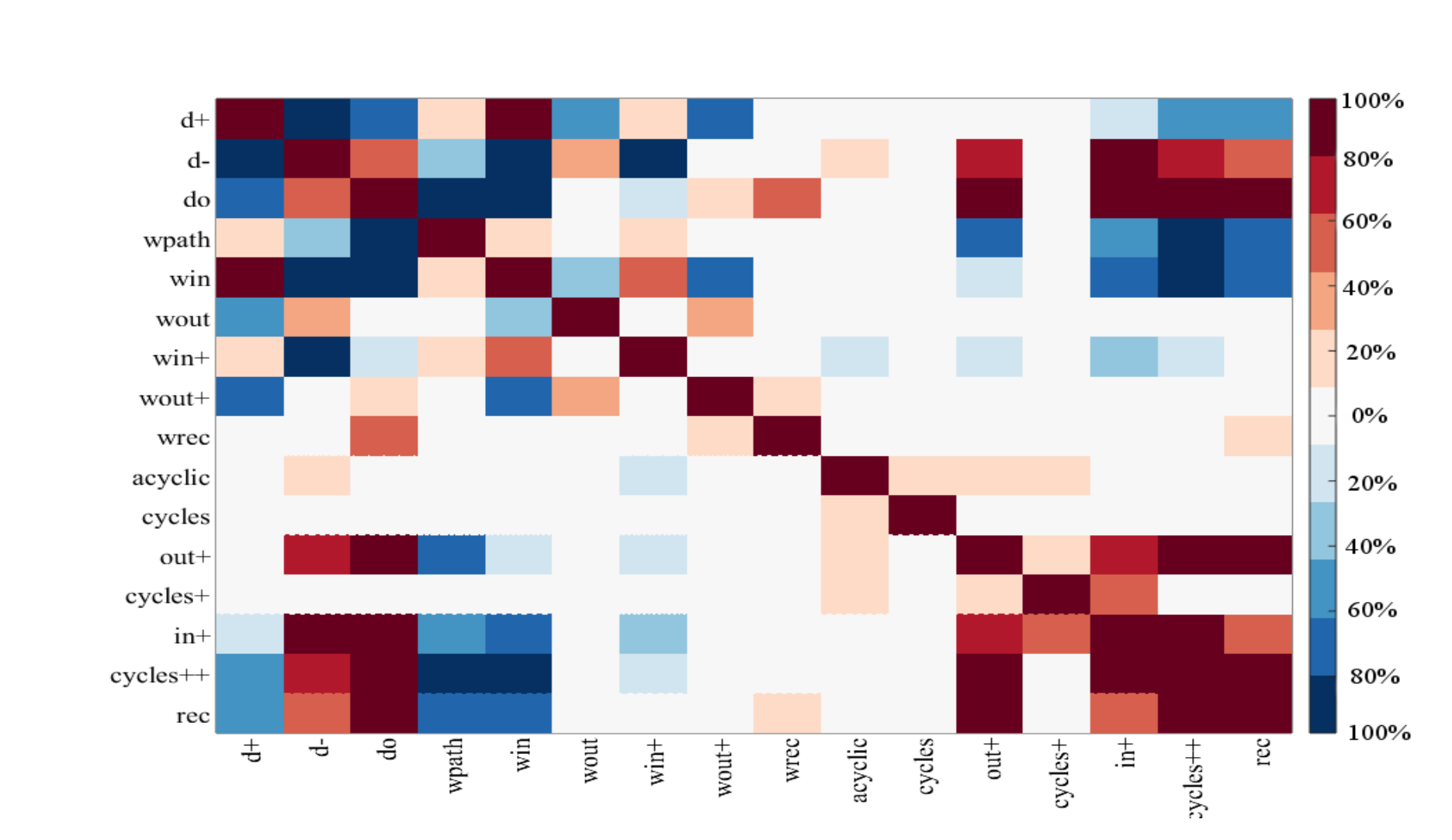}	
\label{fig:2B}
\end{center}
\caption{Graphlet correlation matrices are computed for all 40 healthy subjects.  Percentages of the healthy subjects that have statistically significant are colored.  The correlation is considered significant if the Pearson correlation coefficient is greater than $0.7$ and for anti-correlation is significant if the coefficient is less than $-0.7$. The heat-map indicates that there are many pairs of entries of the signature vector that are significantly correlated or anti-correlated for most of the subjects.}
\end{figure}

We then compute the percentage of subjects for which particular coordinates (similarity metrics) of the signature vectors are significantly correlated ($>0.7$) or anti-correlated ($<-0.7$), Figure 3. We found that for brain effective networks higher values of the number of out-degree is associated (significantly anti-correlated) with lower values of the number if in-degree in 95\% subjects and with the number of reciprocal-degree in 64\% subjects. Therefore, regions which are involved in activations (G-causes) of other regions (higher values of G-causes manifested with higher values of $d^+$) at the same time are not activate (G-effects) by other regions (lower values of G-effects shown in both $d^-$ and $d^\circ$).  On the other hand, the number of in-degree and the number of reciprocal-degree are correlated in 62\% subjects.  Moreover, we found that $W^{(path)}$ is anti-correlated with $T^{(out+)}$ (in 74\% subjects), $T^{(cycle++)}$ (in 85\% subjects) and with $T^{(rec)}$ (in 79\% subjects). The number of in-wedges   $W^{(in)}$ is anti-correlated with $W^{(out+)}$ (in 74\% subjects), $T^{(in+)}$ (in 79\% subjects), ), $T^{(cycle++)}$ (in 85\% subjects) and with $T^{(rec)}$ (in 74\% subjects). While the relationships between degrees, wedge-degrees, and wedge-degrees and triangle-degrees are mostly negative (anti-correlated), we found that relationships between different number of triangles (involving reciprocal edges) are positive (correlated): (1) $T^{(out+)}$ with $T^{(in+)}$ (69\%), $T^{(cycle++)}$ (100\%), $T^{(rec)}$ (97\%); (2) $T^{(cycle+)}$ with $T^{(in+)}$ (59\%); (3) $T^{(in+)}$ with $T^{(cycle++)}$ (100\%) and $T^{(rec)}$ (59\%); (4) $T^{(cycle++)}$ with $T^{(rec)}$ (100\%). However, no significant (positive or negative) relationships are found in which the number of acyclic triangles and the number of cycle triangles are involved.

In conclusion vertex signature vector and graphlet correlation matrix are powerful tools for network analysis.  We found that the graphlet correlation matrix of the brain effective network has positive correlation between various triangle-degrees only if at least one edge is reciprocal. Negative correlation was pronounced when comparing out-degrees with in-degrees and reciprocal-degrees, as well as, when comparing degrees with wedge-degrees. How this finding will further enhance our understanding of brain connectivity will be a subject of our future research activities. 

After submitting this paper we learned that Przulj's group was working on similar problem using different approach and data \cite{przulj2016}.

\clearpage

\begin{center}
\begin{LARGE}
Supplementary Information to the paper \\``Graphlet characteristics in directed networks''\\~\\~\\
\end{LARGE}

Here we give detailed explanations and provide additional examples to the main paper \cite{ourpaper}. 
\end{center}



\section{Graphlet-based similarity}

In undirected graphs, there are eight 2- to 4-node graphlets resulting in 15 orbits \cite{przulj-2} (of which 11 are non-redundant \cite{przulj-2}), that touch a vertex $i$. Since the number of up to 4-mode graphlets in directed networks is large, here we consider only a subset of these orbits starting (originating) from a vertex. More specifically, let $ h_1, ... h_k,$ be nodes belonging to graphlets with $2 \leq k \leq 4$ nodes and define $\alpha_1, \alpha_2, \alpha_3,\alpha_4,\alpha_5,\alpha_6 \in \{+, -, \circ \}$. We enumerate all possible ways in which an orbit can touch the node $h_1$ in these directed graphlets:
\begin{eqnarray*}
O_0(h_1)^{(\alpha_1)} &=& S_{h_1}^{\alpha_1} \\ 
O_1(h_1)^{(\alpha_1,\alpha_2)}  &=& \left\{ h_2,h_3:  h_2 \in S_{h_1}^{\alpha_1}, h_3 \in S_{h_2}^{\alpha_2}  \right\}  \\  
O_2(h_1)^{(\alpha_1,\alpha_2)} &=& \left\{ h_2,h_3: h_2 \in S_{h_1}^{\alpha_1}, h_3 \in S_{h_1}^{\alpha_2}  \right\}  \\   
O_3(h_1)^{(\alpha_1,\alpha_2,\alpha_3)} &=& \left\{ h_2,h_3: h_2 \in S_{h_1}^{\alpha_1}, h_3 \in S_{h_1}^{\alpha_2} \cap S_{h_2}^{\alpha_3} \right\}   \\
O_4(h_1)^{(\alpha_1,\alpha_2,\alpha_3)} &=& \left\{ h_2,h_3,h_4: h_{i+1} \in S_{h_i}^{\alpha_i}, i=1,2,3 \right\}  \\
O_5(h_1)^{(\alpha_1,\alpha_2,\alpha_3)} &=& \left\{ h_2,h_3,h_4: h_2 \in S_{h_1}^{\alpha_1}, h_3 \in S_{h_1}^{\alpha_2}, h_4 \in S_{h_3}^{\alpha_3} \right\}   \\
O_6(h_1)^{(\alpha_1,\alpha_2,\alpha_3)} &=& \left\{h_2,h_3,h_4: h_2 \in S_{h_1}^{\alpha_1}, h_3 \in S_{h_2}^{\alpha_2}, h_4 \in S_{h_2}^{\alpha_3} \right\} \\
O_7(h_1)^{(\alpha_1,\alpha_2,\alpha_3)} &=& \left\{h_2,h_3,h_4: h_i \in S_{h_1}^{\alpha_{i-1}}, i=2,3,4 \right\} \\
O_8(h_1)^{(\alpha_1,\alpha_2,\alpha_3,\alpha_4)} &=& \left\{h_2,h_3,h_4: h_{i+1} \in S_{h_i}^{\alpha_i}, i=1,2,3; h_4 \in S_{h_1}^{\alpha_4} \right\}  \\ 
O_9(h_1)^{(\alpha_1,\alpha_2,\alpha_3,\alpha_4)}  &=& \left\{h_2,h_3,h_4: h_{i+1} \in S_{h_i}^{\alpha_i}, i=1,2,3; h_4 \in S_{h_2}^{\alpha_4} \right\} \\
O_{10}(h_1)^{(\alpha_1,\alpha_2,\alpha_3,\alpha_4)}  &=& \left\{h_2,h_3,h_4: h_{i+1} \in S_{h_i}^{\alpha_i}, i=1,2,3; h_3 \in S_{h_1}^{\alpha_4} \right\} \\
O_{11}(h_1)^{(\alpha_1,\alpha_2,\alpha_3,\alpha_4)}  &=& \left\{h_2,h_3,h_4: h_{i} \in S_{h_1}^{\alpha_{i-1}}, i=2,3,4; h_4 \in S_{h_3}^{\alpha_4} \right\}  \\
O_{12}(h_1)^{(\alpha_1,\alpha_2,\alpha_3,\alpha_4,\alpha_5)}  &=& \left\{h_2,h_3,h_4: h_{i+1} \in S_{h_i}^{\alpha_i}, i=1,2; h_4 \in S_{h_i}^{\alpha_{i+2}}, i=1,2,3 \right\} \\
O_{13}(h_1)^{(\alpha_1,\alpha_2,\alpha_3,\alpha_4,\alpha_5)}  &=& \left\{h_2,h_3,h_4: h_2 \in S_{h_1}^{\alpha_1}, h_3 \in S_{h_1}^{\alpha_2}; h_4 \in S_{h_i}^{\alpha_{i+2}}, i=1,2,3 \right\} \\
O_{14}(h_1)^{(\alpha_1,\alpha_2,\alpha_3,\alpha_4,\alpha_5,\alpha_6)}  &=& \left\{h_2,h_3,h_4: h_2 \in S_{h_1}^{\alpha_1}, h_{3} \in S_{h_1}^{\alpha_2} \cap S_{h_2}^{\alpha_3},  h_{4} \in S_{h_1}^{\alpha_4} \cap S_{h_2}^{\alpha_5} \cap S_{h_3}^{\alpha_6} \right\} 
\end{eqnarray*}
resulting in a total of $ 1695 = 3 + 2\times 3^2 + 5\times 3^3 + 4 \times 3^4 + 2 \times 3^5 + 3^6$ combinations.

Considering only 3-node graphlets and focusing only on orbits which \textit{originate} at node $i$, we define the following 39 quantities for a node $i$ in the graph:
\begin{eqnarray*}
d_i^\alpha & =  & \left | S_i^\alpha  \right| \\
T_i^{(\alpha, \beta, \gamma)} &=&  \sum_{j \in S_i^\gamma} \left| S_i^\alpha \cap S_j^\beta \right| \\
{W}_i^{(\alpha, \beta)} & = & \sum_{j \neq i} \left| S_i^\alpha \cap S_j^\beta \right|  - \sum_\gamma T_i^{(\alpha, \beta, \gamma)}  
\end{eqnarray*}
These 39 quantities can be aggregated into 16 quantities by grouping wedges in 6 wedge isomorphic classes and triangles in 7 triangle isomorphic classes, resulting in: 
\begin{itemize}
\item 3 degrees: $d_i^+$, $d_i^-$, and $d_i^\circ$;

\item 6 wedge-degrees:  
\begin{eqnarray*}
W_i^{(path)} & = & {W}_i^{(+,-)} + {W}_i^{(-,+)}  \\
W_i^{(in)} & = & {W}_i^{(+,+)}  \\
W_i^{(out)} & = & {W}_i^{(-,-)}  \\
W_i^{(in+)} & = & {W}_i^{(+,\circ)} + {W}_i^{(\circ,+)}  \\
W_i^{(out+)} & = & {W}_i^{(-,\circ)} + {W}_i^{(\circ,-)}  \\
W_i^{(rec)} & = & {W}_i^{(\circ,\circ)};    
\end{eqnarray*}

\item 7 triangle-degrees:
\begin{eqnarray*}
  T_i^{(acyclic)} &=&   T_i^{(+,-,+)} + T_i^{(+,+,-)} + T_i^{(+,+,+)} + T_i^{(-,-,-)} + T_i^{(-,-,+)} +  T_i^{(-,+,-)} \\ 
  T_i^{(cicles)} &=&  T_i^{(+,-,-)} +  T_i^{(-,+,+)} \\
  T_i^{(out+)} &=& T_i^{(+,\circ, +)} + T_i^{(-,-, \circ )} +T_i^{(\circ, +, -)} \\
 T_i^{(cycles+)} &=& T_i^{(+,-, \circ)} + T_i^{(+, \circ, -)} +T_i^{(-,+, \circ)} + T_i^{(-, \circ, +)} + T_i^{(\circ, -, -)} +T_i^{(\circ, +, +)}  \\
 T_i^{(in+)} &=& T_i^{(+, +, \circ)} +T_i^{(-,\circ, -)} +T_i^{(\circ, -,+)} \\
  T_i^{(cycle++)} &=&  T_i^{(+,\circ,\circ)} +T_i^{(-,\circ, \circ)} + T_i^{(\circ, -, \circ)} +T_i^{(\circ, +, \circ)} + T_i^{(\circ, \circ, -)} +  T_i^{(\circ, \circ, +)} \\ 
 T_i^{(rec)} &=&  T_i^{(\circ, \circ, \circ)}. 
\end{eqnarray*}

\end{itemize}

The quantities degrees, wedge-degrees, and triangle-degrees could be normalized in different ways. For example, one can consider normalized signature vector defined as: 
\begin{eqnarray*}
\bar{F}_i &=& \left[  \frac{d_i^+}{ A_i}, \frac{d_i^-}{A_i}, \frac{d_i^\circ}{A_i}, \frac{W_i^{(path)}}{B_i}, \ldots, \frac{W_i^{(rec)}}{B_i}, \frac{T_i^{(acyclic )}}{C_i}, \ldots, \frac{T_i^{(rec)}}{C_i} \right]^T
\end{eqnarray*}
where 
\begin{eqnarray*}
A_i &=& d_i^+ + d_i^- + d_i^\circ  \\ 
B_i & = & W_i^{(path)} +  W_i^{(in)} + W_i^{(out)} + W_i^{(in+)} + W_i^{(out+)} + W_i^{(rec)}, \\
C_i & = & T_i^{(acyclic)} + T_i^{(cycles)} + T_i^{(out+)} + T_i^{(cycles+)} + T_i^{(in+)} + T_i^{(cycles++)} + T_i^{(rec)}
\end{eqnarray*}
By flipping the direction of edges between all connected pairs of nodes with probability one-third (for out-,in-, and reciprocal-edges),  which keeps the undirected connectivity and the edge density unchanged,  information about the direction of edges will be destroyed and normalized signature vectors should be close to $[\frac{1}{3}, \frac{1}{3}, \frac{1}{3}, \frac{1}{6} \ldots \frac{1}{6}, \frac{1}{7}, \ldots \frac{1}{7}]^T$. This could be further exploited by studying normalized signature vectors for different real-world networks and their distances to random directed network with no direction-generated structure (such graph could be generated by considering arbitrary undirected connected graph and then transforming it to directed graph by assigning to each edge of the graph a direction (out, in, or reciprocal) with probability 1/3).

The number of triangles (but also wedges) could also be normalized as follows.   
$t_i(\alpha, \beta, \gamma) $ is the normalized number of  $(\alpha, \beta, \gamma)$ triangles associated with the actor $i$: 
\begin{eqnarray*}
t_i(\alpha, \beta, \gamma) & =& \frac{\sum_{j \in S_i^{\gamma}} | S_i^{\alpha} \cap S_j^{\beta} |}{\sum_{j \neq i} | S_i^{\alpha} \cap S_j^{\beta} |} 
= 
\frac{ \sum_j x_\gamma \sum_h x_\alpha x_\beta}{ \sum_j \sum_h x_\alpha x_\beta}   
\end{eqnarray*} 
%
$ \sum_{j \neq i} | S_i^\alpha \cap  S_j^\beta|  = \sum_{j \neq i} \sum_h x_\alpha x_\beta$ is the number  of all $(\alpha, \beta)$ wedges starting at $i$. Thus, 
$t_i(\alpha,\beta,\gamma)$ indicates the ratio of $(\alpha, \beta)$ wedges that are  $(\alpha, \beta, \gamma)$ triangles. This is a generalization of the concept of clustering coefficient of a graph to clustering coefficient of $(\alpha, \beta, \gamma)$ triangle associated to the node $i$:     
$$
t_i = \frac{ \mbox{ Number of } (\alpha, \beta, \gamma) \mbox{ triangles}}{ \mbox{Number of } (\alpha, \beta) \mbox{ wedges}} 
$$

\section{Effective brain networks}

\subsection{Data description}

The effective brain networks were extracted from the rs - fMRI images of a group of healthy subjects and were taken from the Beijing Enhanced data set for multimodal brain images \cite{BeijingEnhanced}. In this study, a total of 40 healthy subjects were used, aged from 18 to 26 years. The fMRI scans were taken during resting state of the subject, with a SIMENS Trio Tim scanner (whose strength of magnetics field of 3 Tesla, and EPI sequence). The following parameters of the EPI sequence were used: repetition (TR) / time echo (TE) = 2000 / 30 ms, flip angle = 90, field of view (FOV) = $200 \times 200 mm^2$, voxel dimension = $3 \times 3 \times 3 mm^3$, number of slices = 240. For preprocessing the fMRI data the DPARSF toolbox \cite{Chao-Gan} was used. First, the images of each subject were corrected for slice timing and realigned (motion corrected). Then the images were normalized using the MNI EPI template with affine registration followed by nonlinear transformation. After, smoothing was applied using a Gaussian Kernel of 4mm Full Width at Half Maximum and the signal was detrended to remove any noise that may have remained from the previous steps. Finally, the data were filtered to preserve low frequency fluctuations ($0.01 - 0.08$ Hz). The effective connectivity for each subject were reconstructed using Granger causality coefficient between the time series of 116 regions of interest (ROIs). After obtaining effective connectivity for each subject, the week connections from the effective connectivity network were pruned using threshold value, independently for each subject using. Using a pruning threshold must meet the following two criteria \cite{Papo, Bassett, Meunier} : (i) 99 percent of the vertices (regions) are connected after pruning and (ii) the minimum degree of each vertex is at least $2 \times \ln(N)$, where $N$ is the number of nodes in the network according to Network based statatistic toolbox \cite{Zalesky}.

\subsection{Graphlet correlation matrix}

Here graphlet correlation matrices are provided for all 40 healthy subjects (Figure 1 to Figure 7).

\begin{figure*}[htp]
\centering

\begin{minipage}{.5\textwidth}
\centering
\subfloat[Subject 1]{%
 \includegraphics[trim=3cm 9cm 4cm 8cm,clip,width=\textwidth]{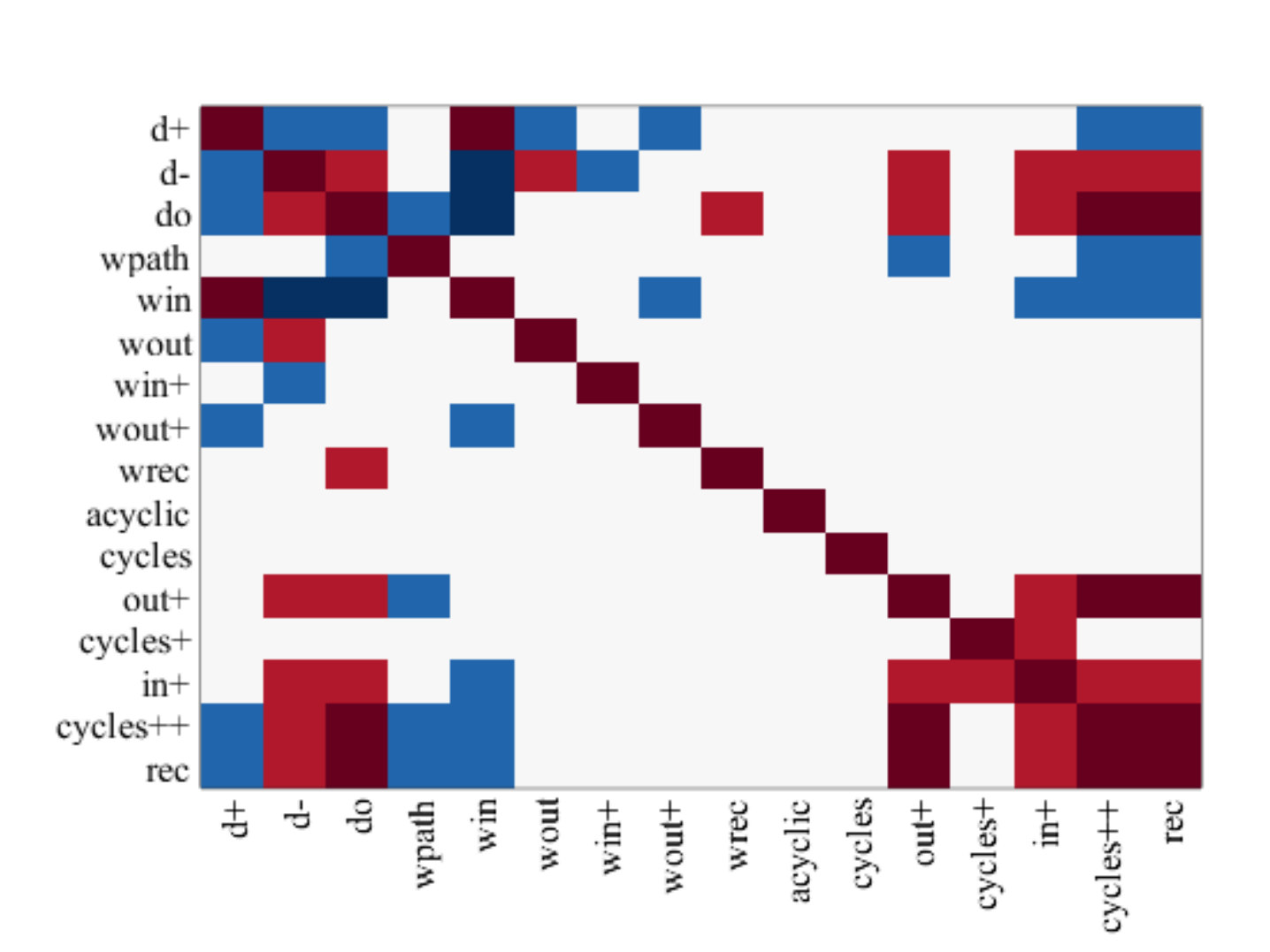}%
}

\subfloat[Subject 2]{%
\includegraphics[trim=3cm 9cm 4cm 8cm, clip, width=\textwidth]{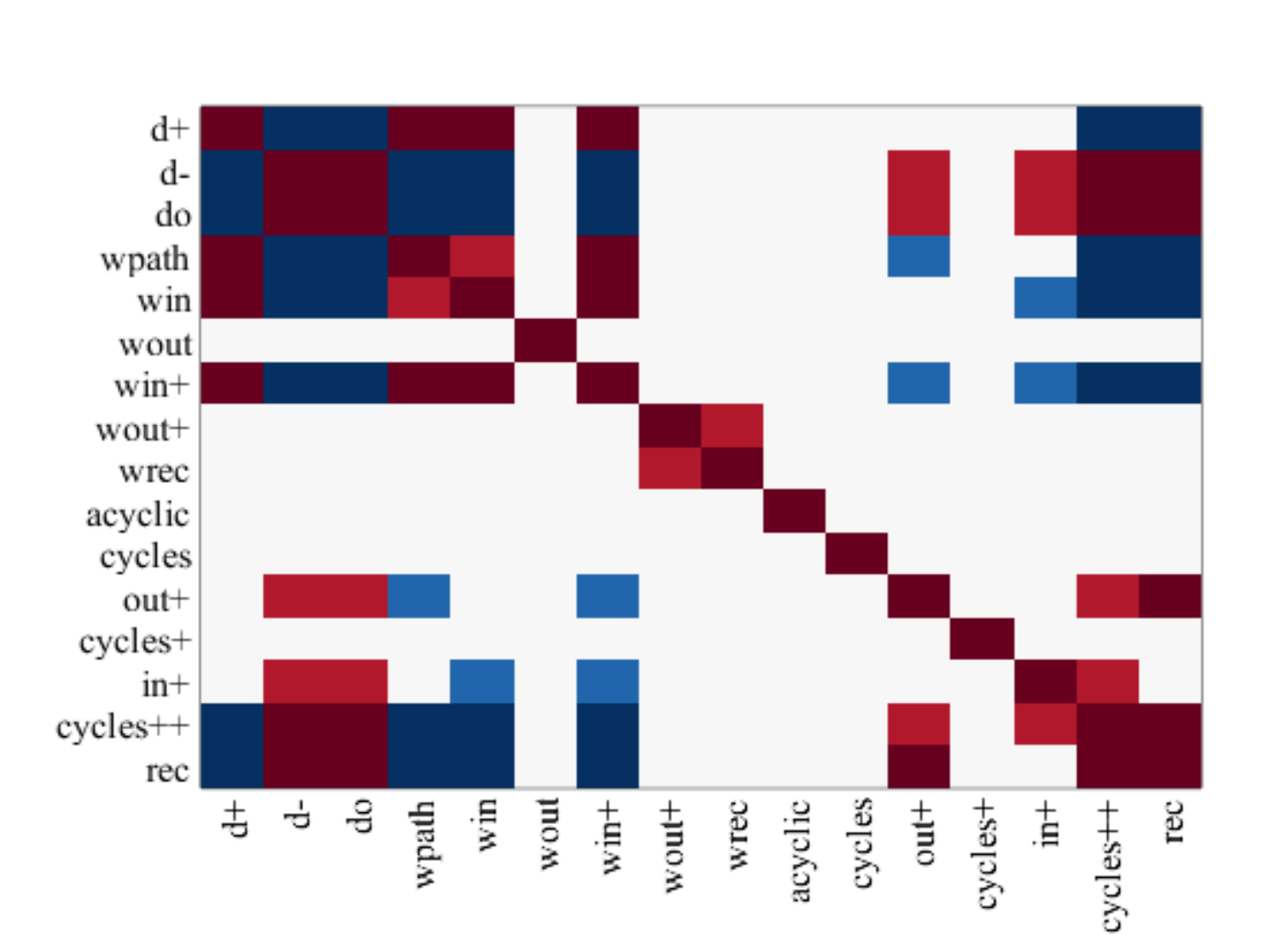}	
}%

\subfloat[Subject 3]{%
\includegraphics[trim=3cm 9cm 4cm 8cm, clip, width=\textwidth]{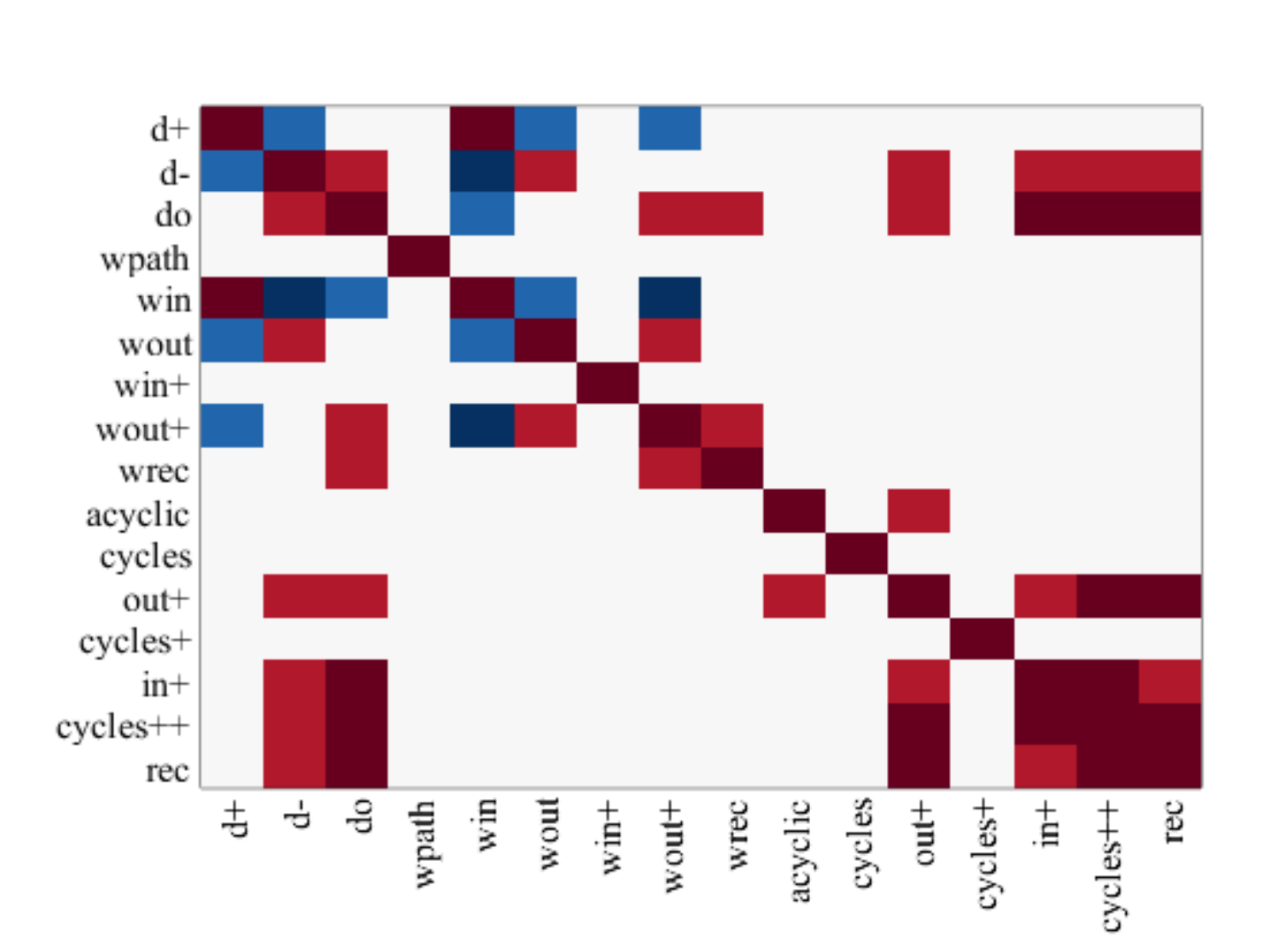}	
}%

\end{minipage}%
\begin{minipage}{.5\textwidth}
\centering
\subfloat[Subject 4]{%
\includegraphics[trim=3cm 9cm 4cm 8cm,clip,width=\textwidth]{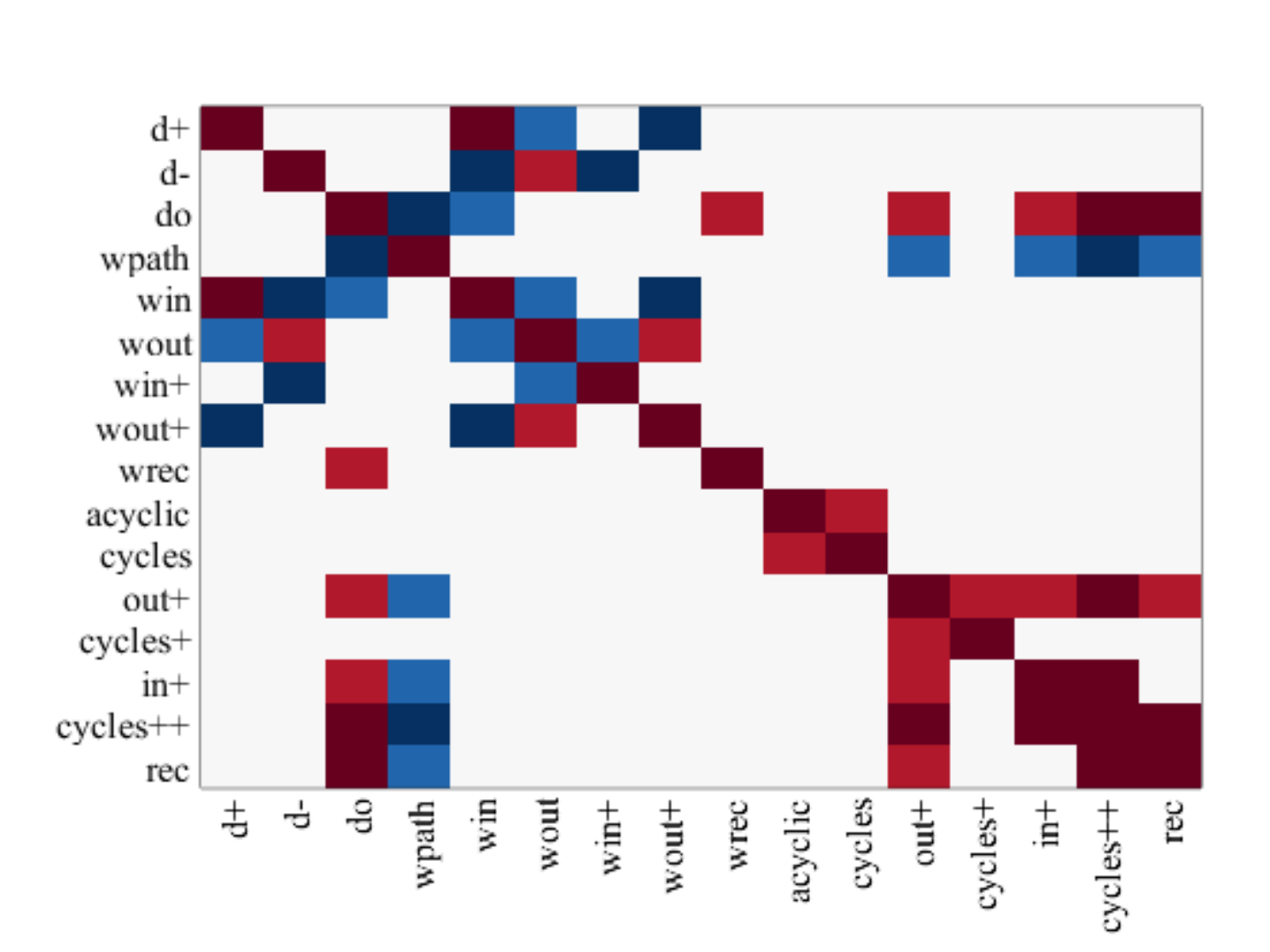}%
}%

\subfloat[Subject 5]{%
\includegraphics[trim=3cm 9cm 4cm 8cm, clip, width=\textwidth]{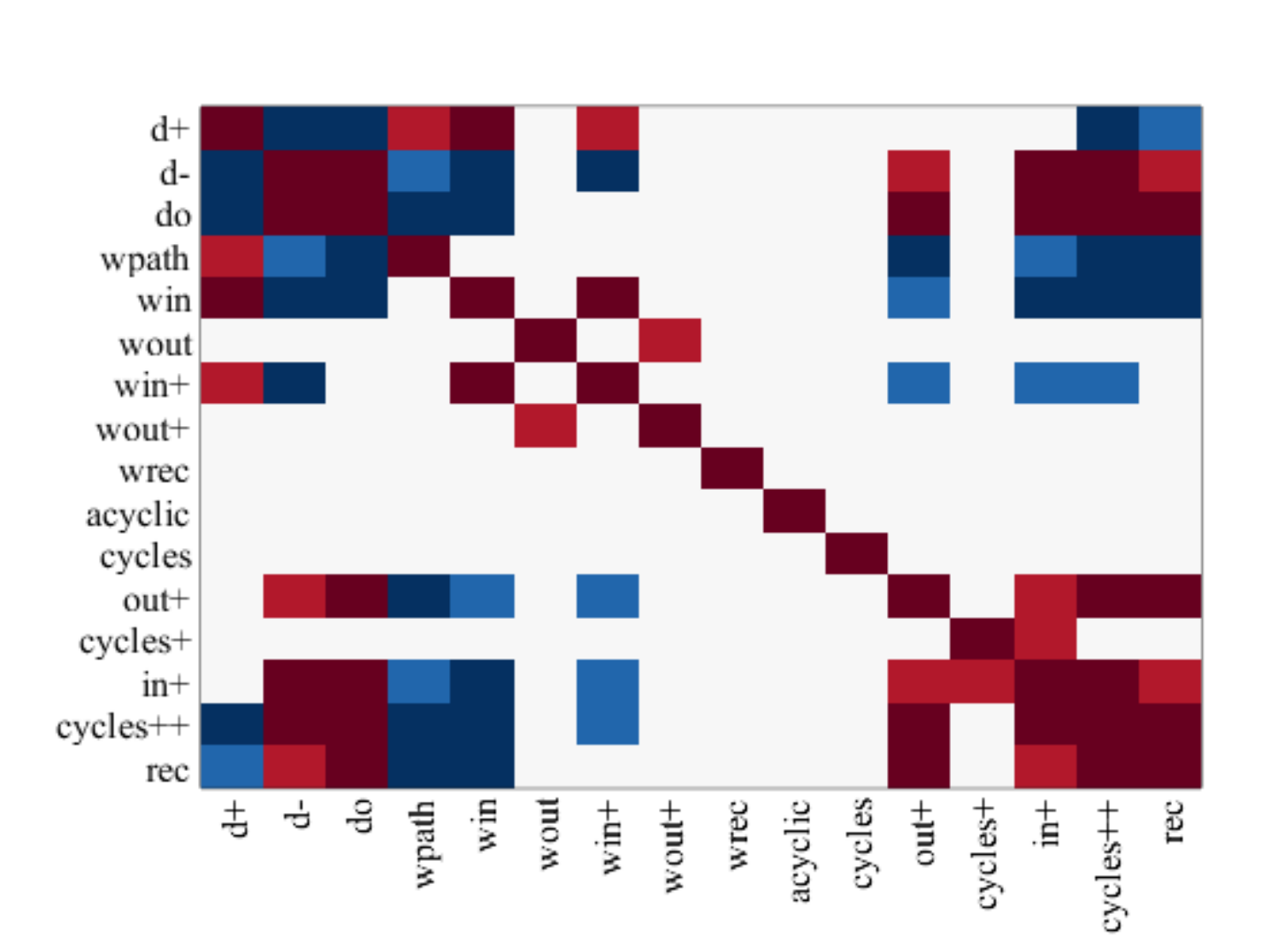}	
}%

\subfloat[Subject 6]{%
\includegraphics[trim=3cm 9cm 4cm 8cm, clip, width=\textwidth]{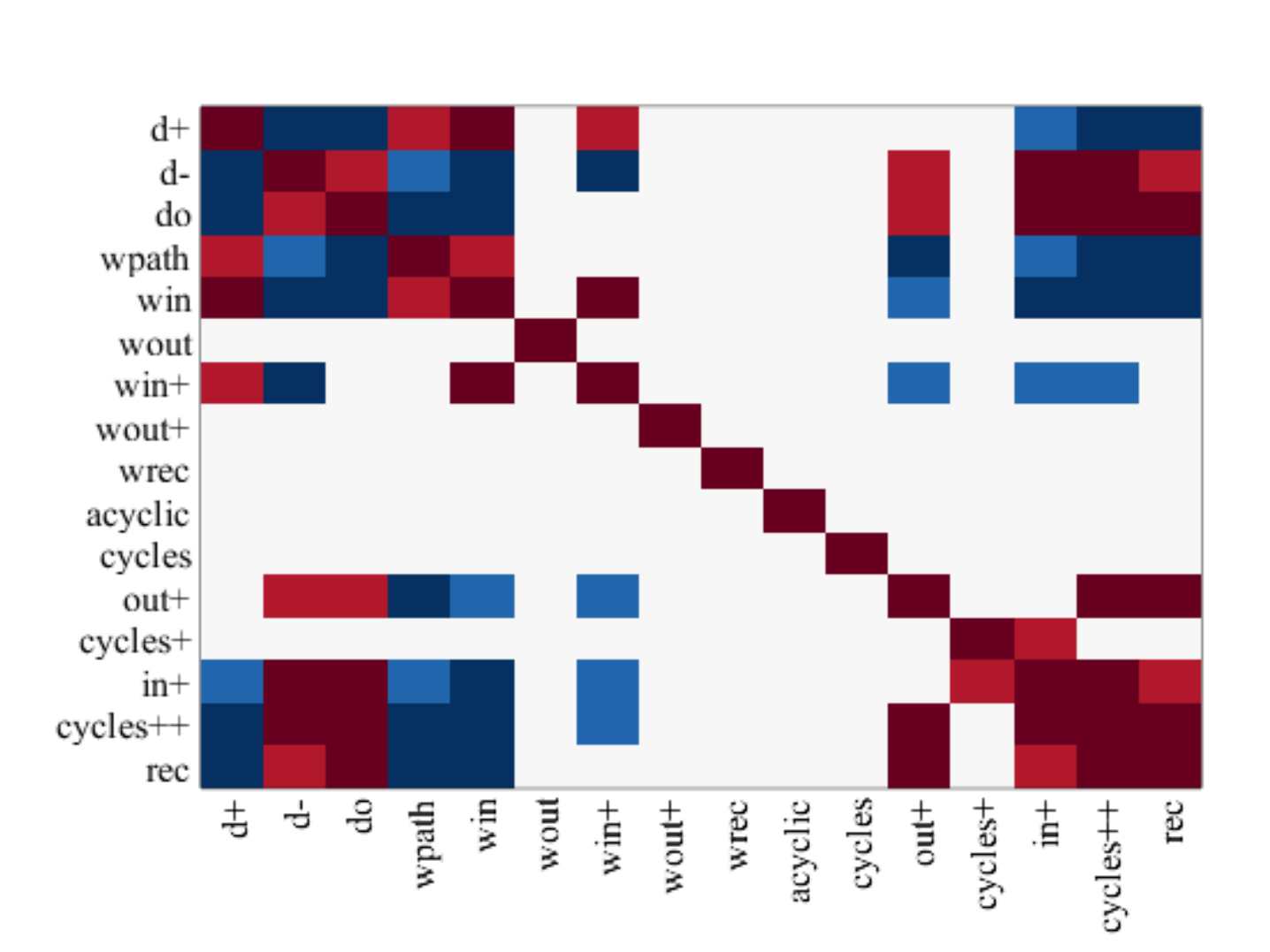}	
}%
\end{minipage}

\caption{Graphlet correlation matrices for subjects 1 to 6}
\end{figure*}

\begin{figure*}[htp]
\centering

\begin{minipage}{.5\textwidth}
\centering
\subfloat[Subject 7]{%
 \includegraphics[trim=3cm 9cm 4cm 8cm,clip,width=\textwidth]{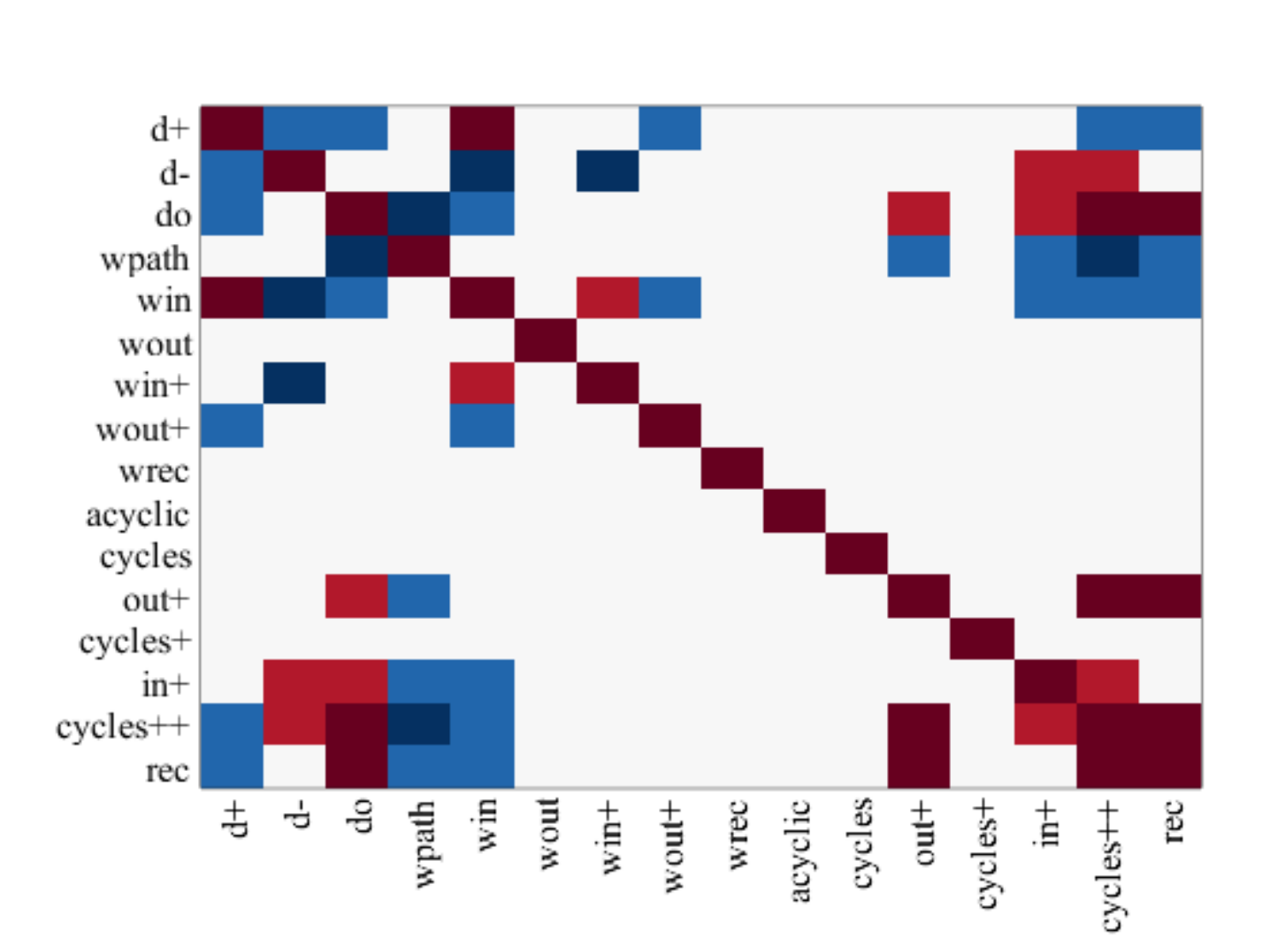}%
}

\subfloat[Subject 8]{%
\includegraphics[trim=3cm 9cm 4cm 8cm, clip, width=\textwidth]{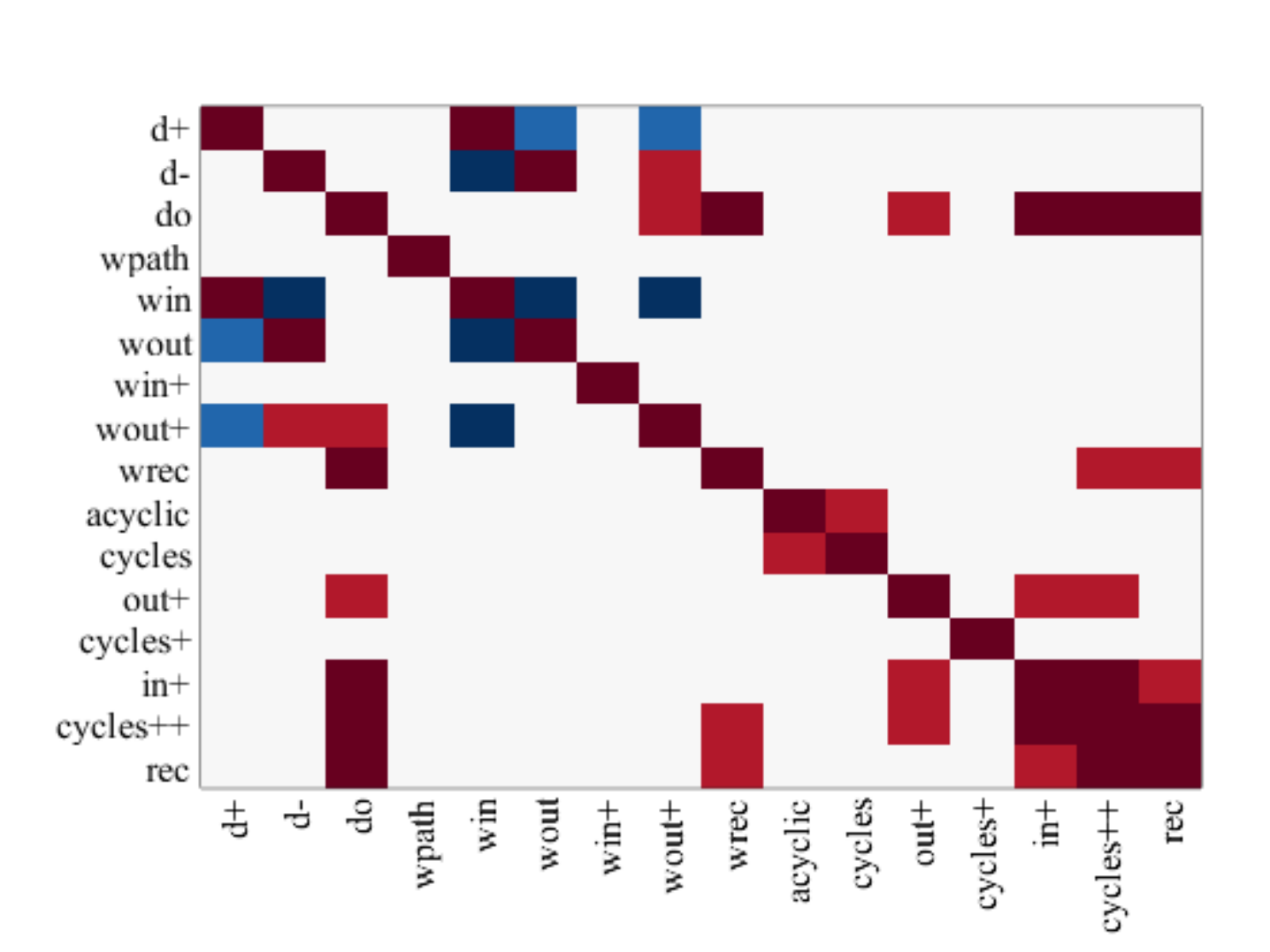}	
}%

\subfloat[Subject 9]{%
\includegraphics[trim=3cm 9cm 4cm 8cm, clip, width=\textwidth]{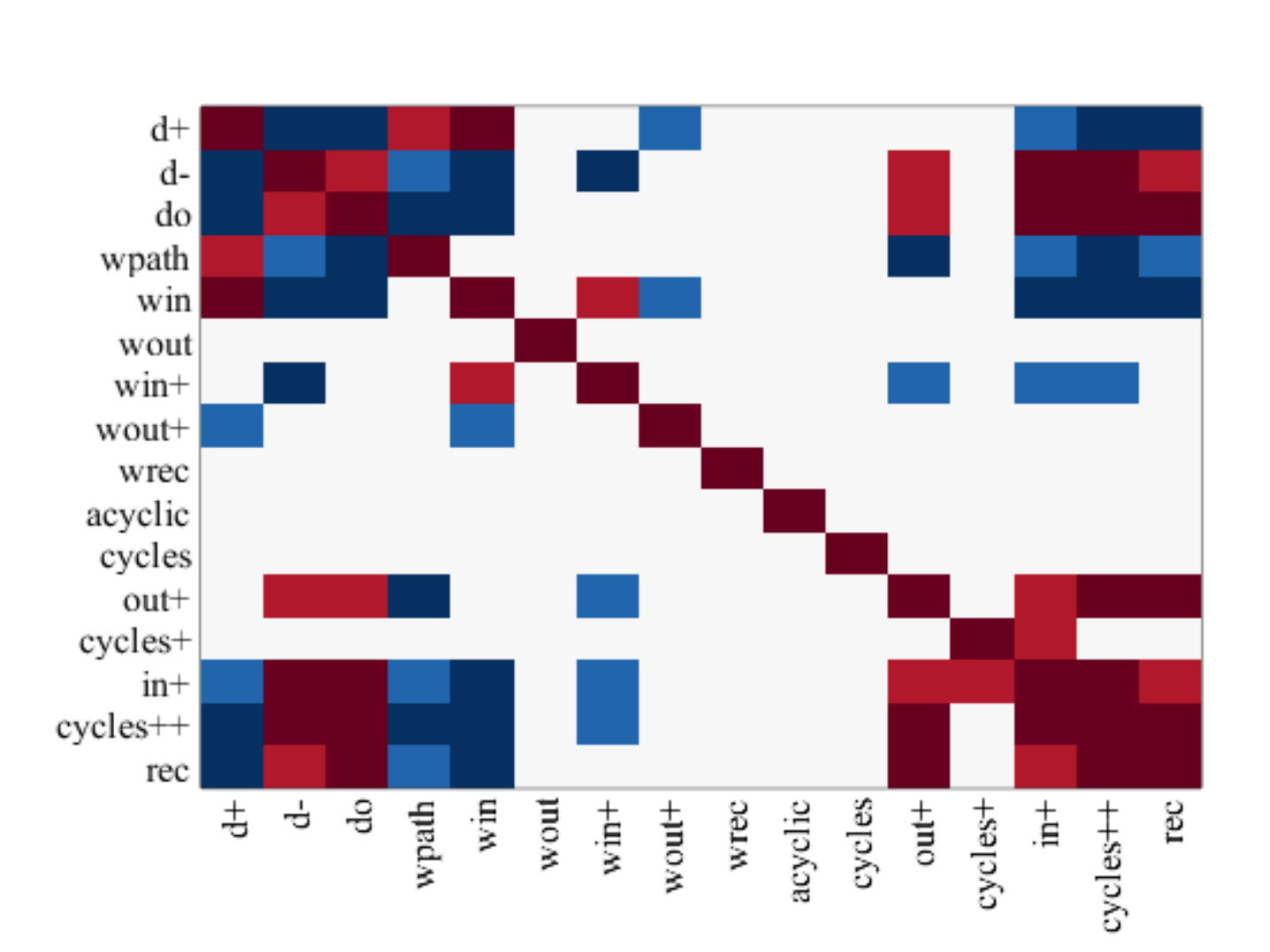}	
}%

\end{minipage}%
\begin{minipage}{.5\textwidth}
\centering
\subfloat[Subject 10]{%
\includegraphics[trim=3cm 9cm 4cm 8cm,clip,width=\textwidth]{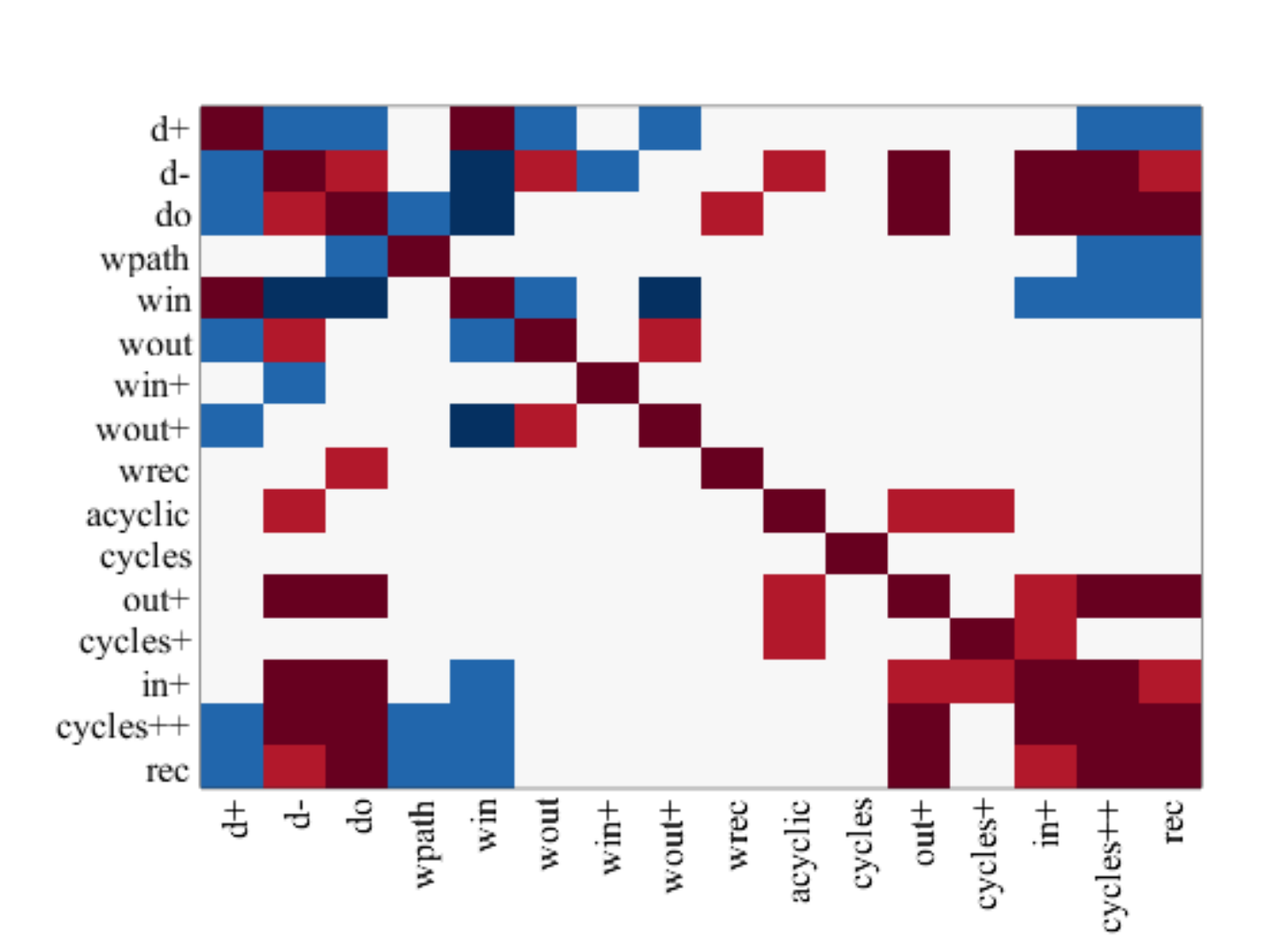}%
}%

\subfloat[Panient 11]{%
\includegraphics[trim=3cm 9cm 4cm 8cm, clip, width=\textwidth]{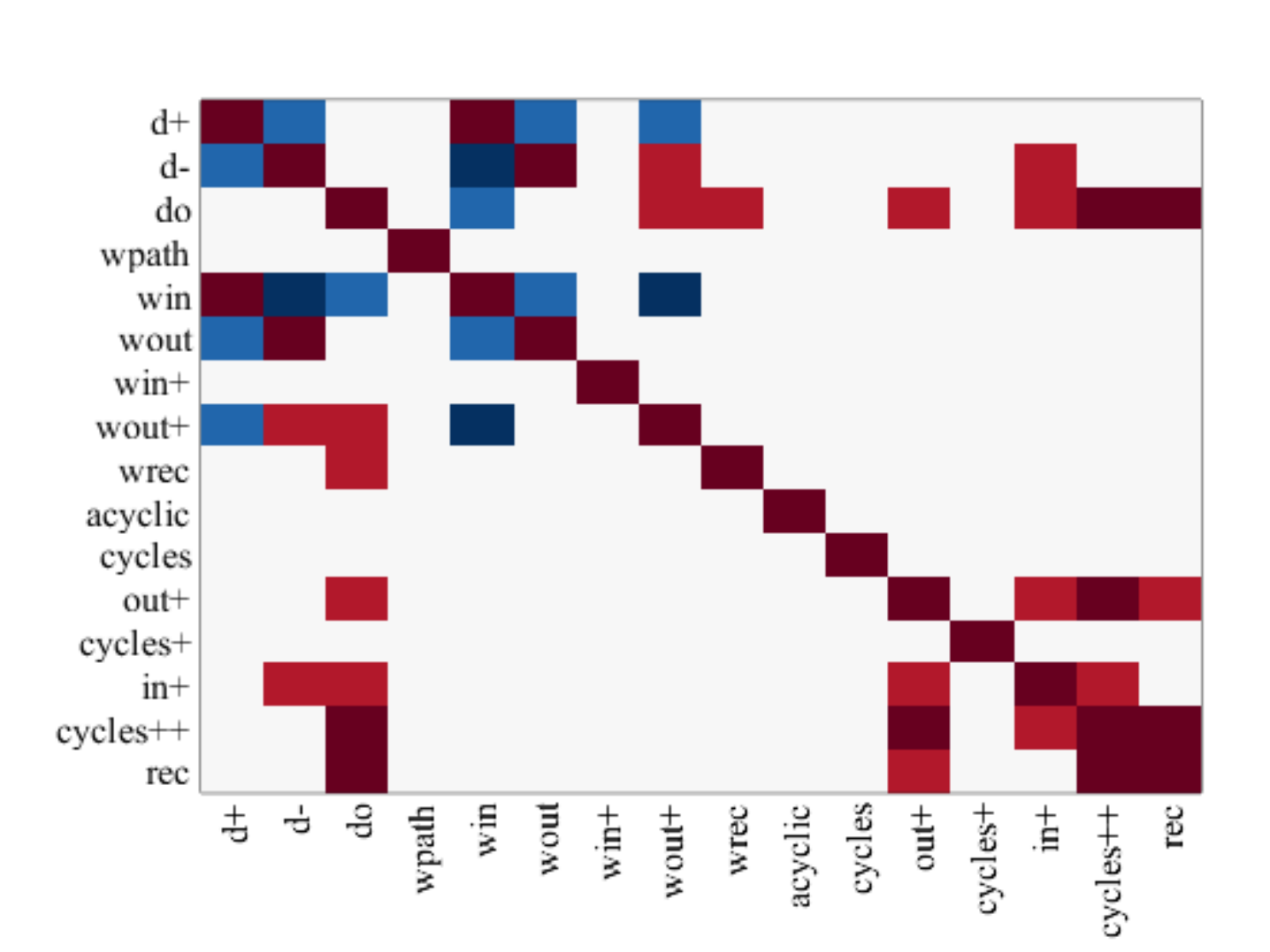}	
}%

\subfloat[Subject 12]{%
\includegraphics[trim=3cm 9cm 4cm 8cm, clip, width=\textwidth]{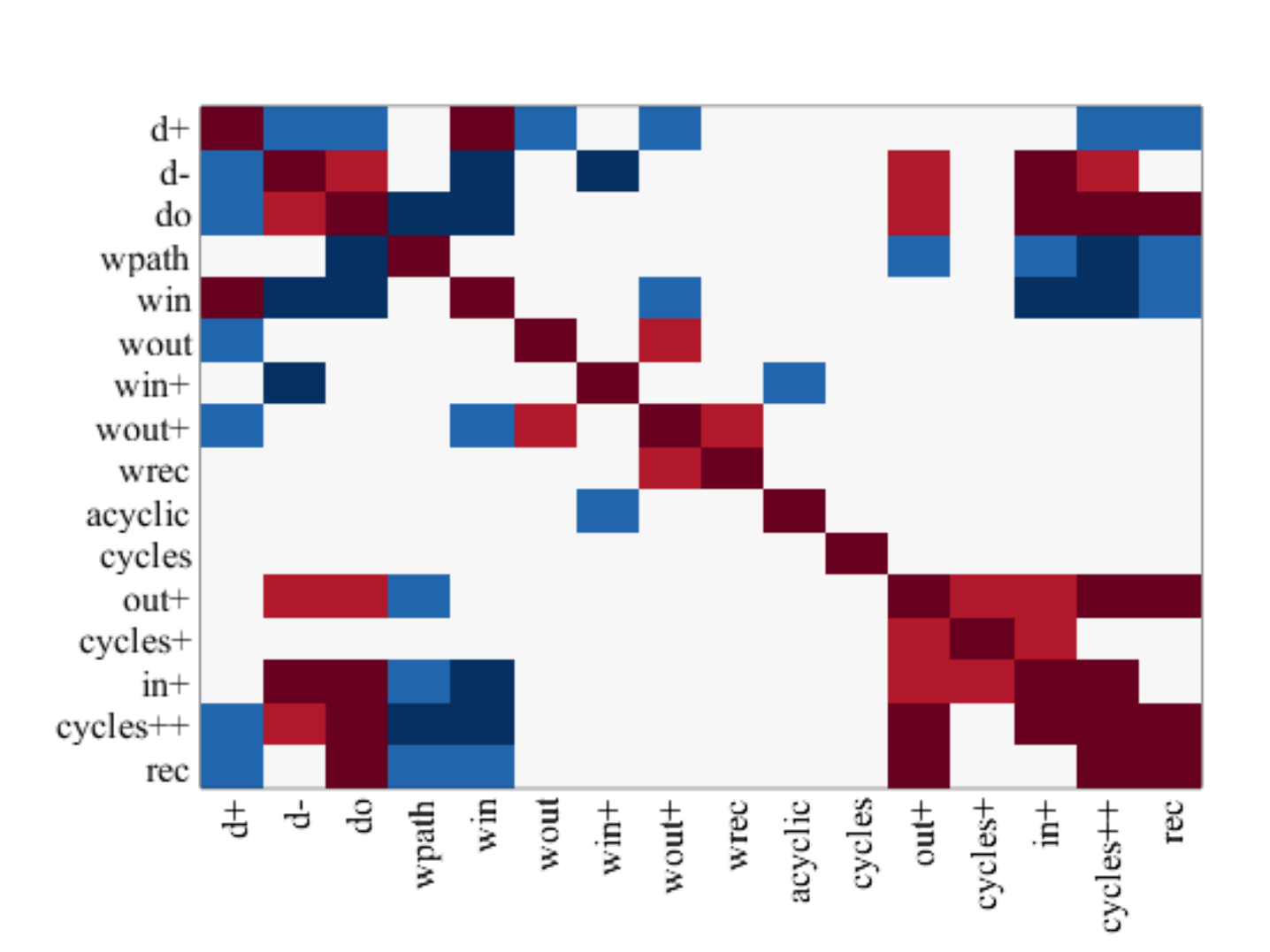}	
}%
\end{minipage}

\caption{Graphlet correlation matrices for subjects 7 to 12}
\end{figure*}

\begin{figure*}[htp]
\centering

\begin{minipage}{.5\textwidth}
\centering
\subfloat[Subject 13]{%
 \includegraphics[trim=3cm 9cm 4cm 8cm,clip,width=\textwidth]{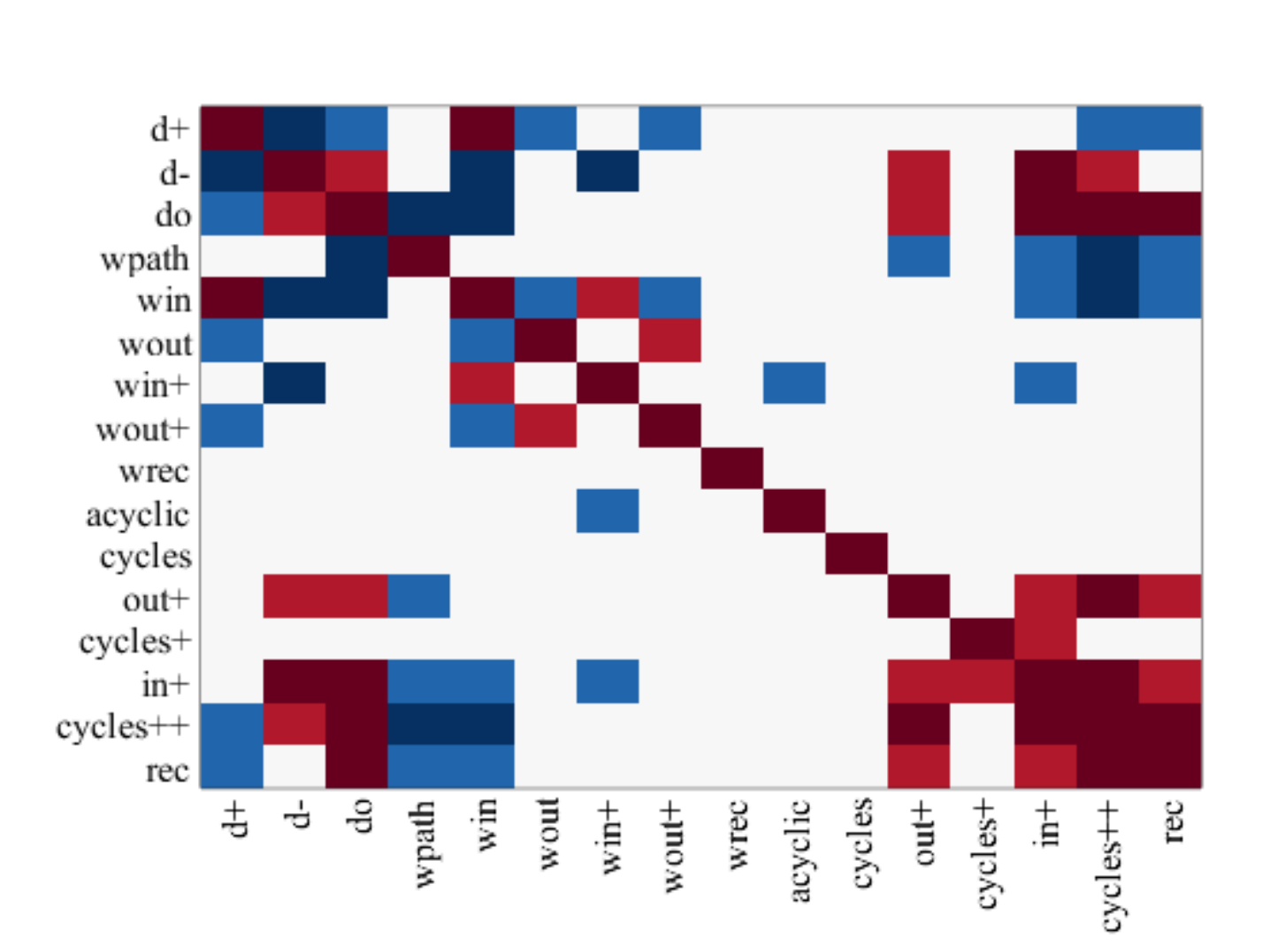}%
}

\subfloat[Subject 14]{%
\includegraphics[trim=3cm 9cm 4cm 8cm, clip, width=\textwidth]{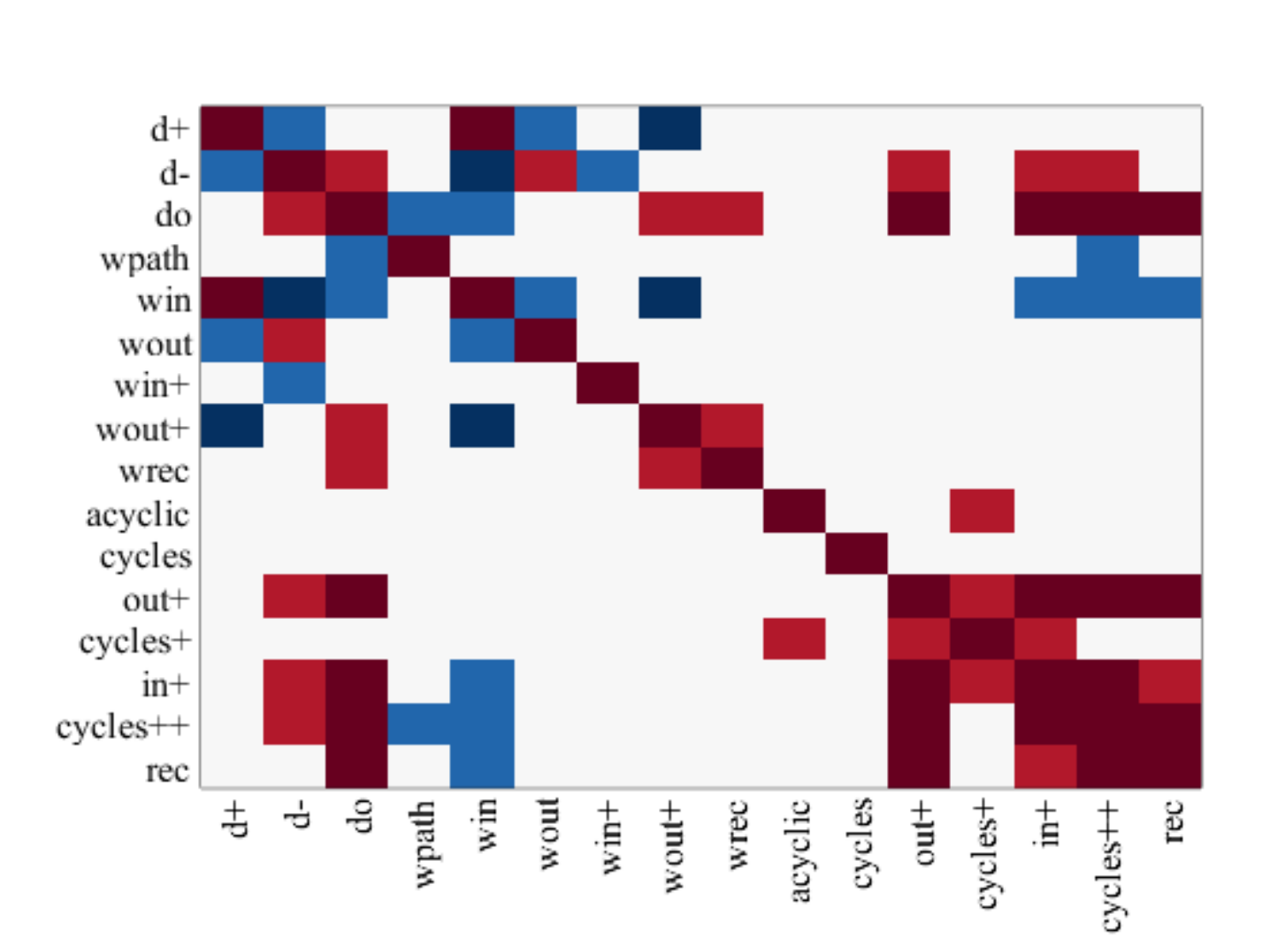}	
}%

\subfloat[Subject 15]{%
\includegraphics[trim=3cm 9cm 4cm 8cm, clip, width=\textwidth]{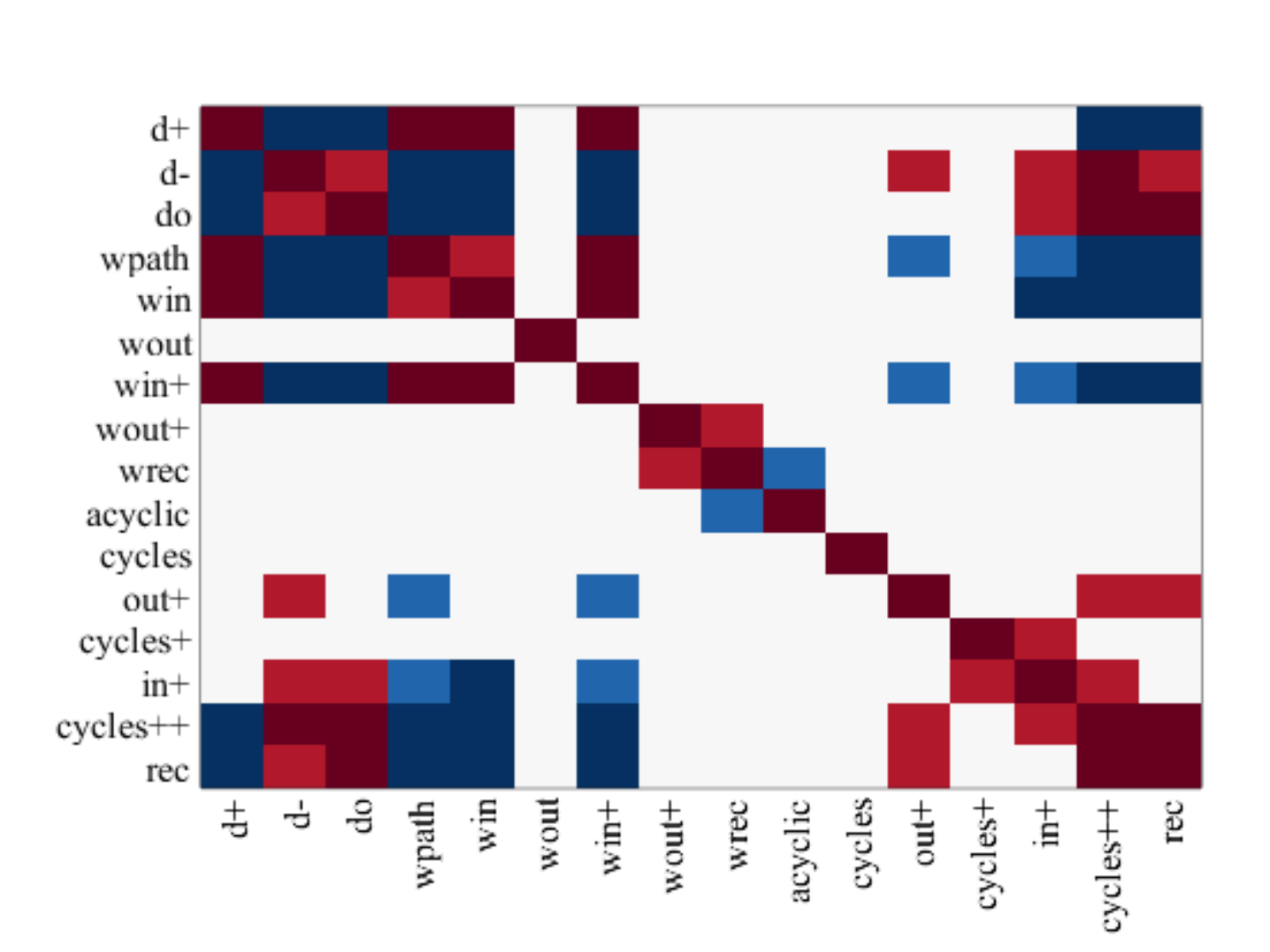}	
}%

\end{minipage}%
\begin{minipage}{.5\textwidth}
\centering
\subfloat[Subject 16]{%
\includegraphics[trim=3cm 9cm 4cm 8cm,clip,width=\textwidth]{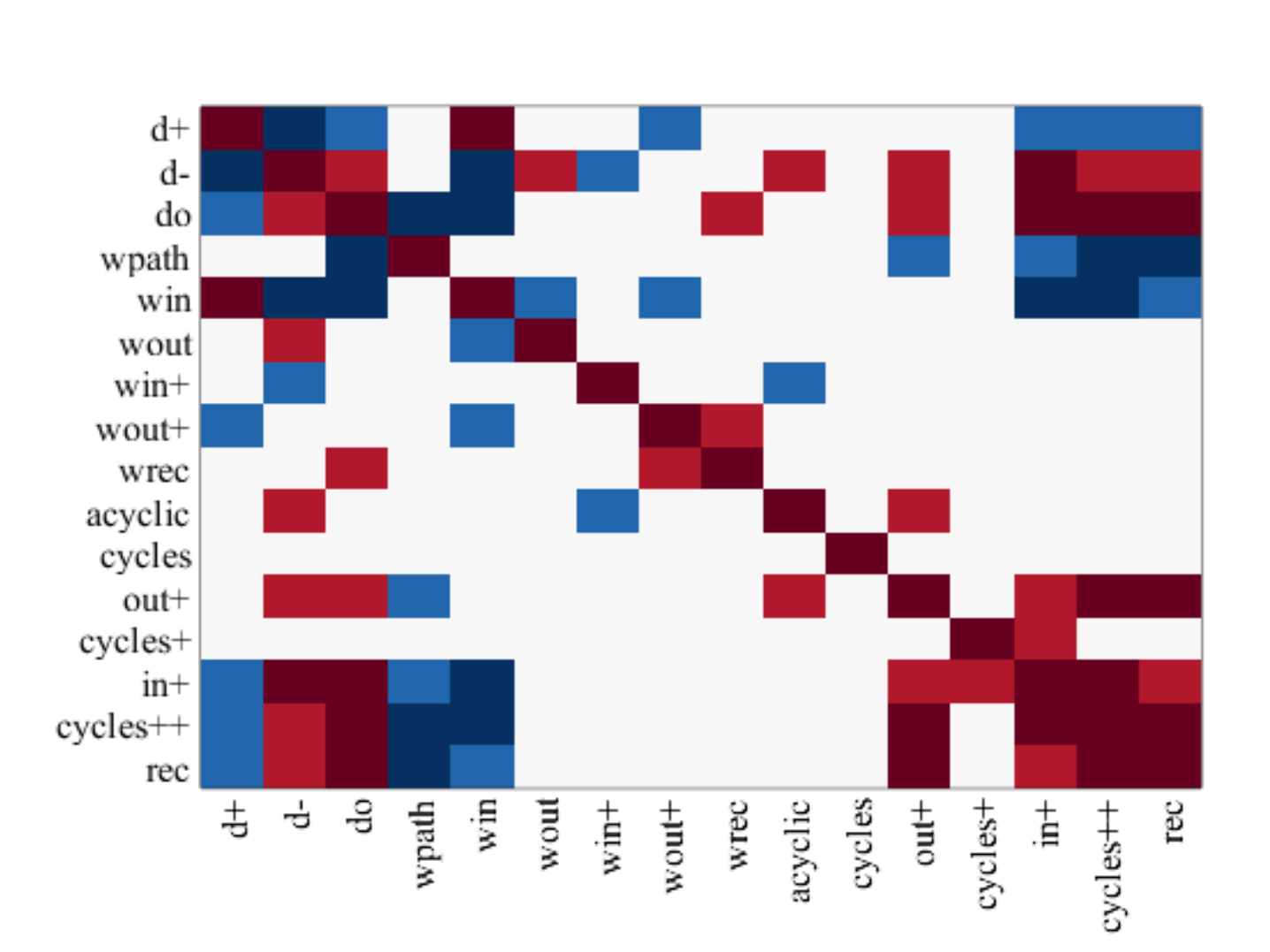}%
}%

\subfloat[Panient 17]{%
\includegraphics[trim=3cm 9cm 4cm 8cm, clip, width=\textwidth]{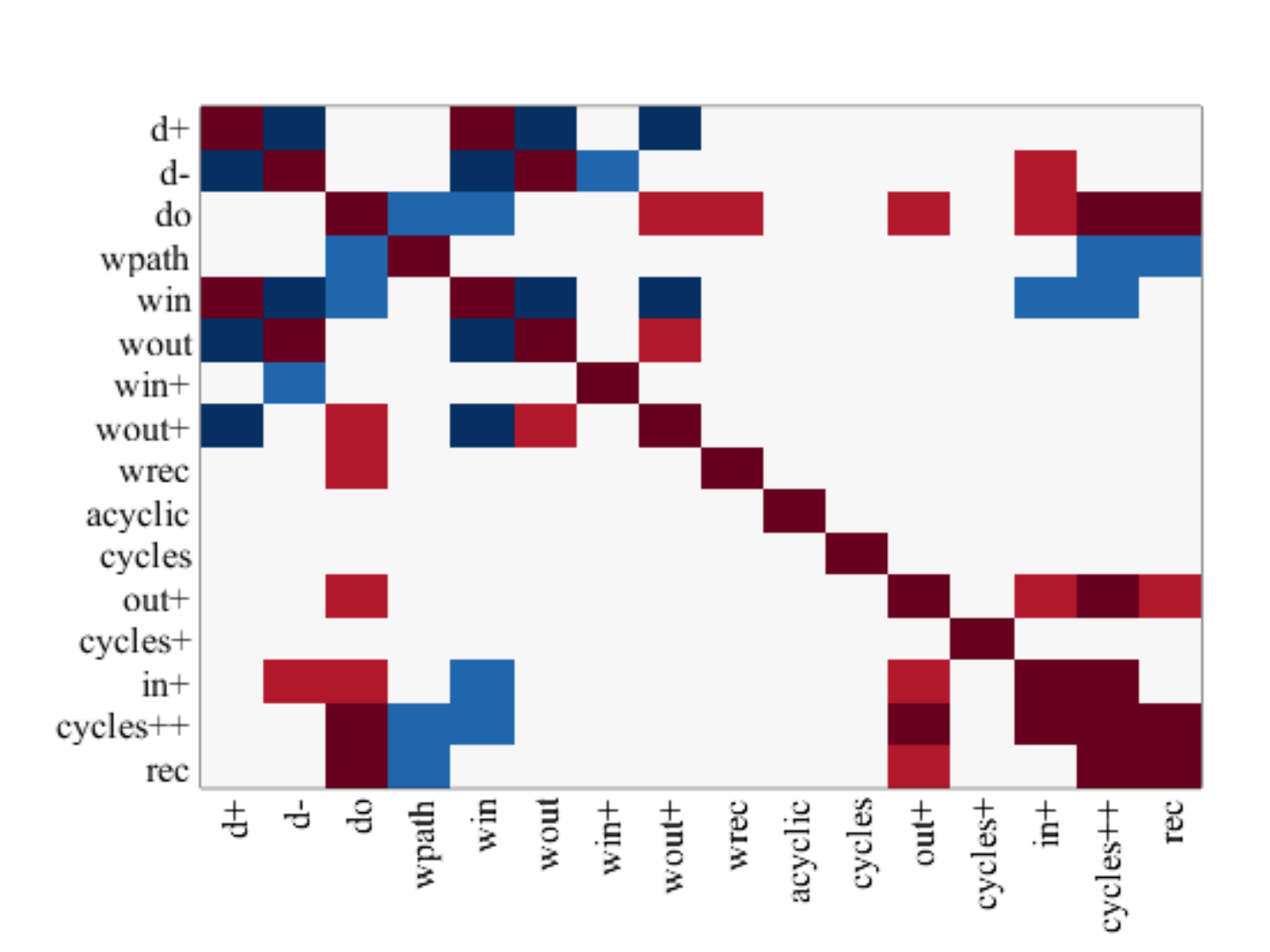}	
}%

\subfloat[Subject 18]{%
\includegraphics[trim=3cm 9cm 4cm 8cm, clip, width=\textwidth]{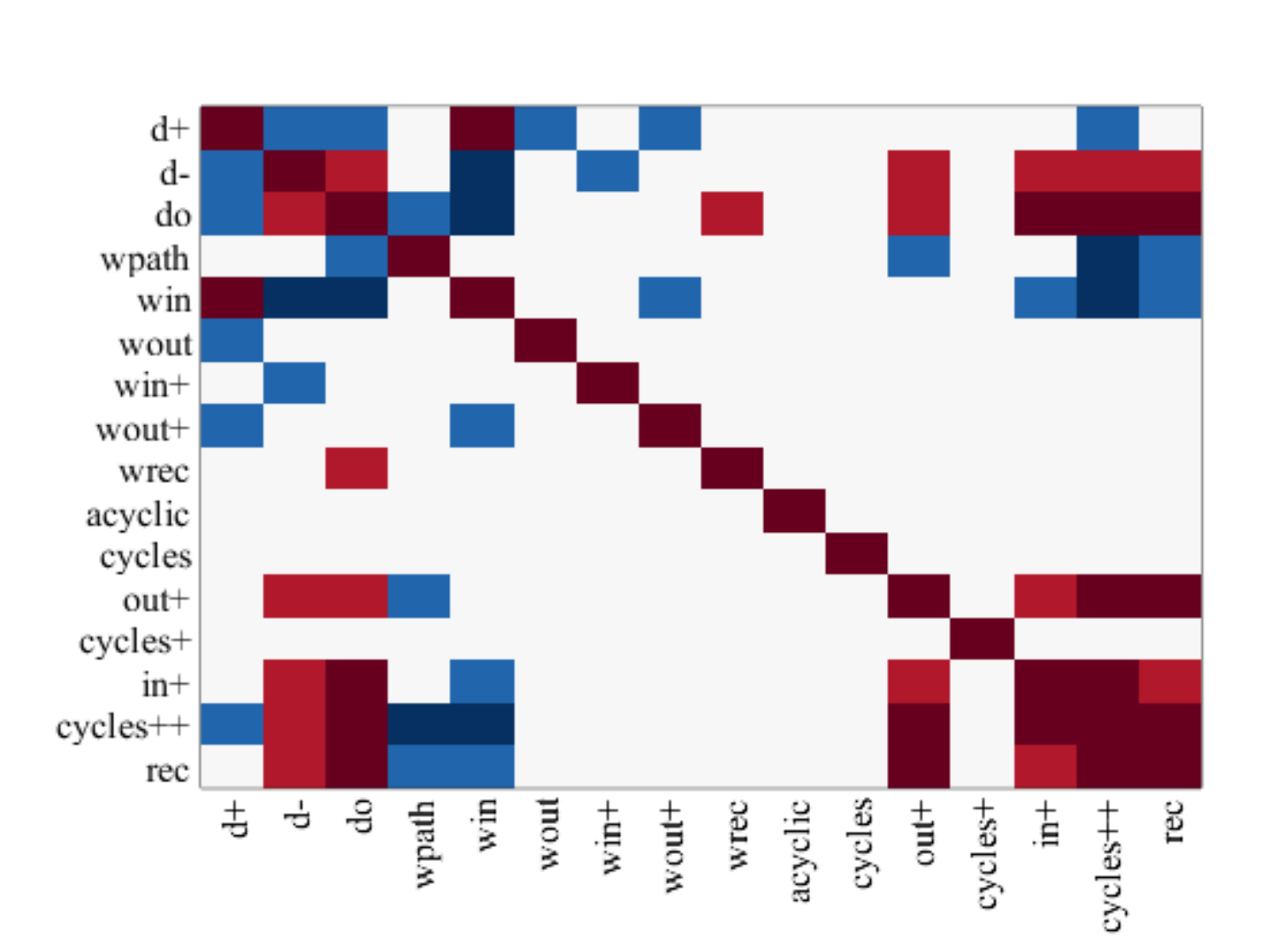}	
}%
\end{minipage}

\caption{Graphlet correlation matrices for subjects 13 to 18}
\end{figure*}

\begin{figure*}[htp]
\centering

\begin{minipage}{.5\textwidth}
\centering
\subfloat[Subject 19]{%
 \includegraphics[trim=3cm 9cm 4cm 8cm,clip,width=\textwidth]{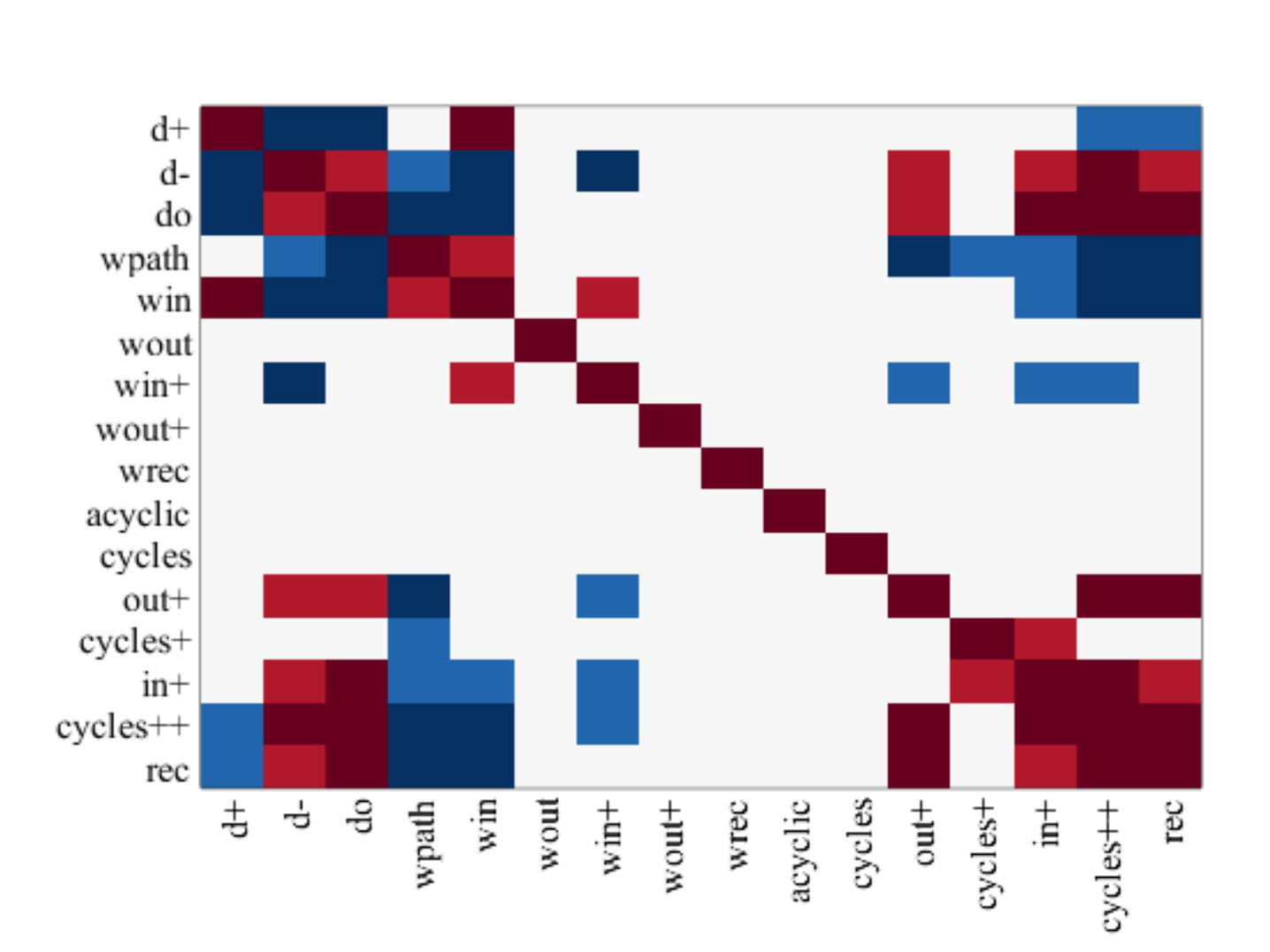}%
}

\subfloat[Subject 20]{%
\includegraphics[trim=3cm 9cm 4cm 8cm, clip, width=\textwidth]{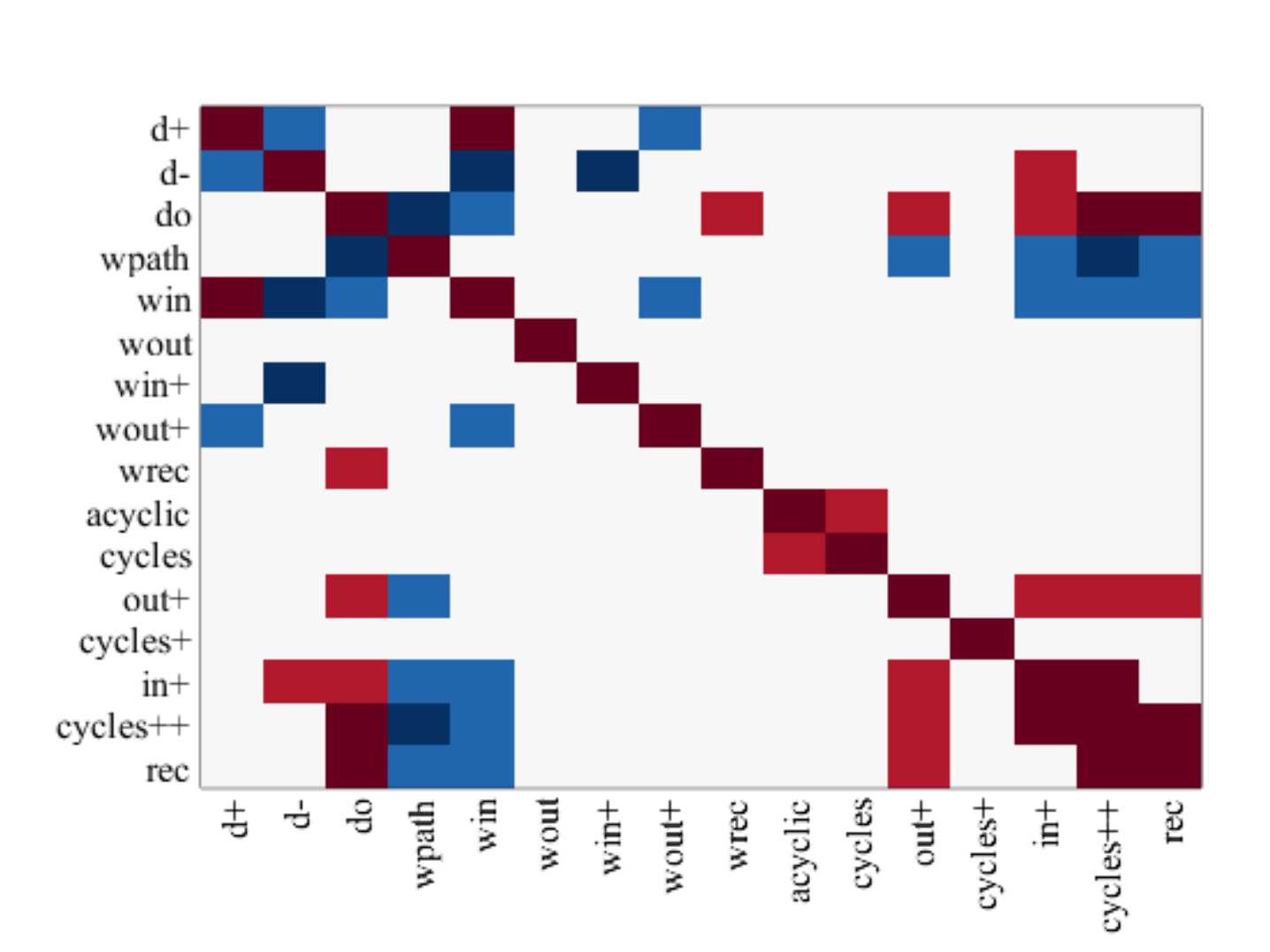}	
}%

\subfloat[Subject 22]{%
\includegraphics[trim=3cm 9cm 4cm 8cm, clip, width=\textwidth]{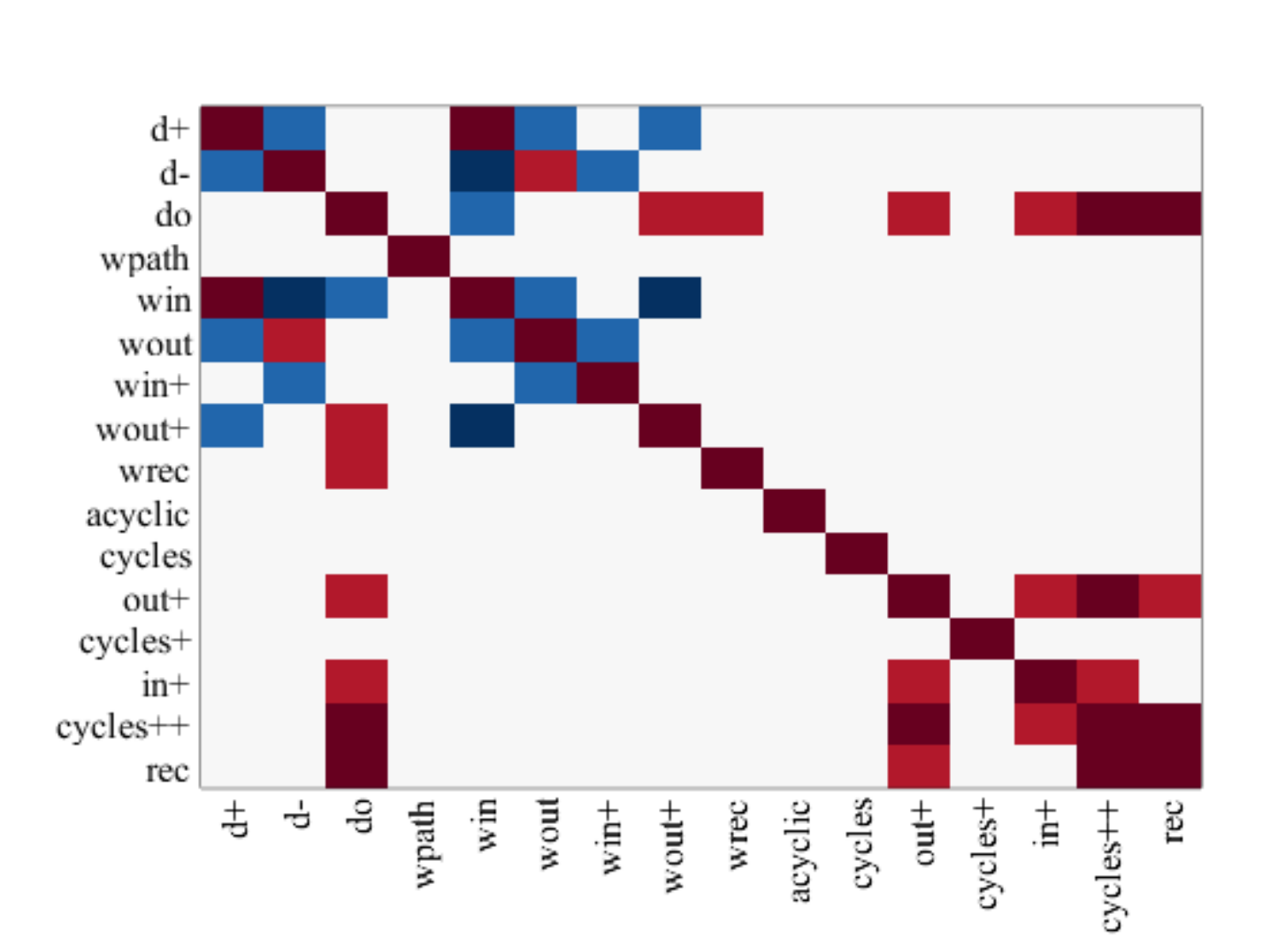}	
}%

\end{minipage}%
\begin{minipage}{.5\textwidth}
\centering
\subfloat[Subject 22]{%
\includegraphics[trim=3cm 9cm 4cm 8cm,clip,width=\textwidth]{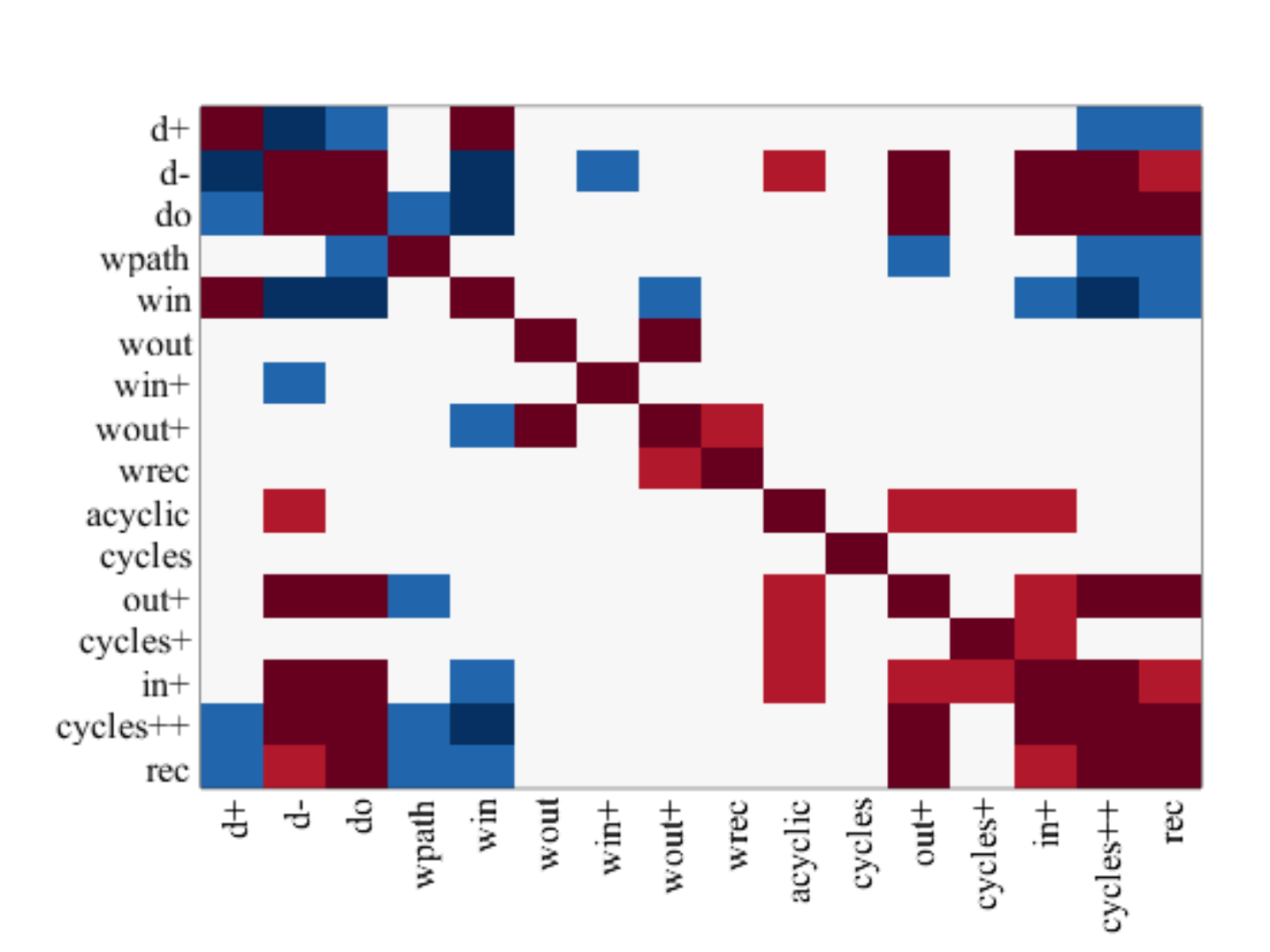}%
}%

\subfloat[Panient 23]{%
\includegraphics[trim=3cm 9cm 4cm 8cm, clip, width=\textwidth]{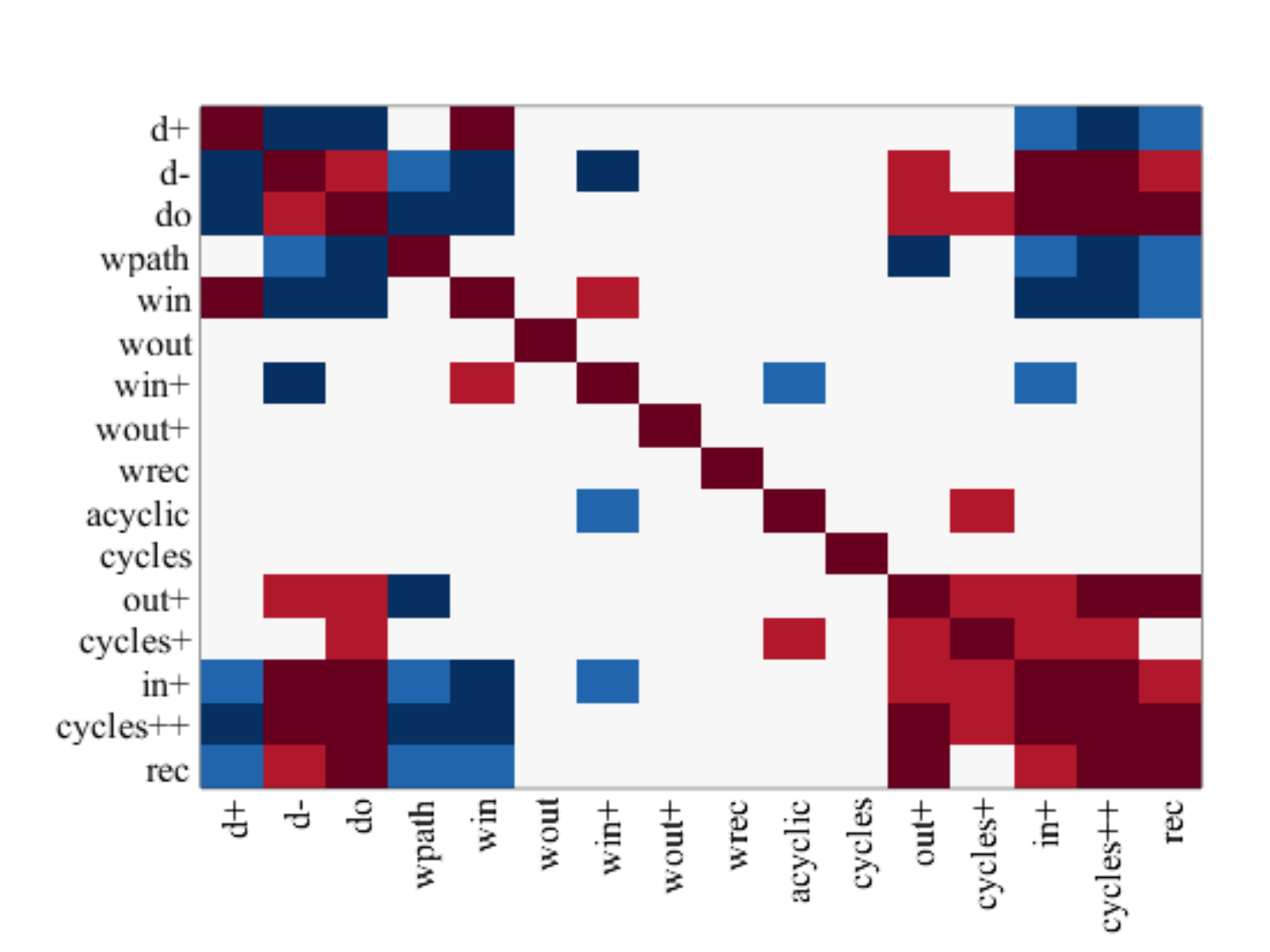}	
}%

\subfloat[Subject 24]{%
\includegraphics[trim=3cm 9cm 4cm 8cm, clip, width=\textwidth]{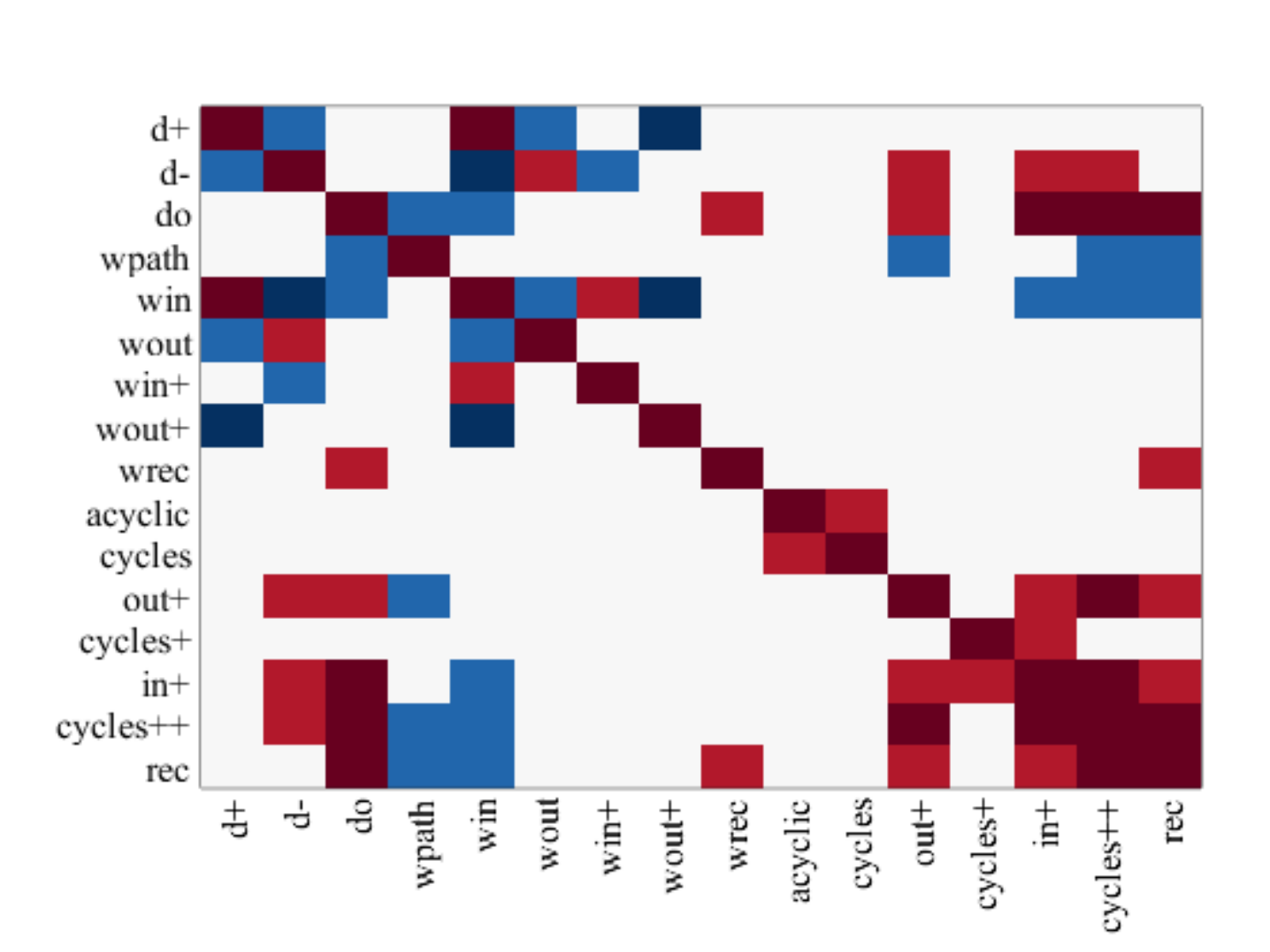}	
}%
\end{minipage}

\caption{Graphlet correlation matrices for subjects 19 to 24}
\end{figure*}

\begin{figure*}[htp]
\centering

\begin{minipage}{.5\textwidth}
\centering
\subfloat[Subject 25]{%
 \includegraphics[trim=3cm 9cm 4cm 8cm,clip,width=\textwidth]{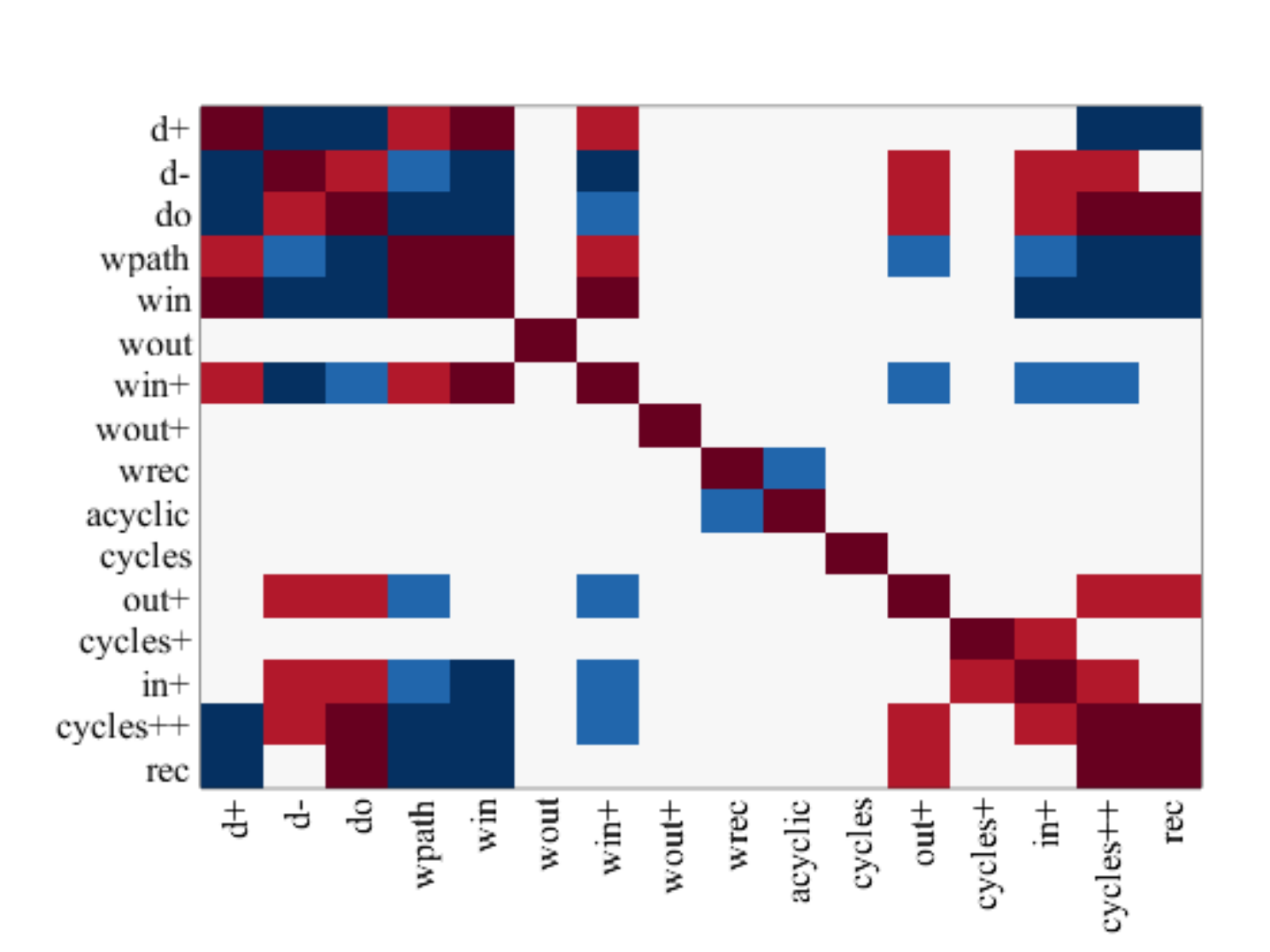}%
}

\subfloat[Subject 26]{%
\includegraphics[trim=3cm 9cm 4cm 8cm, clip, width=\textwidth]{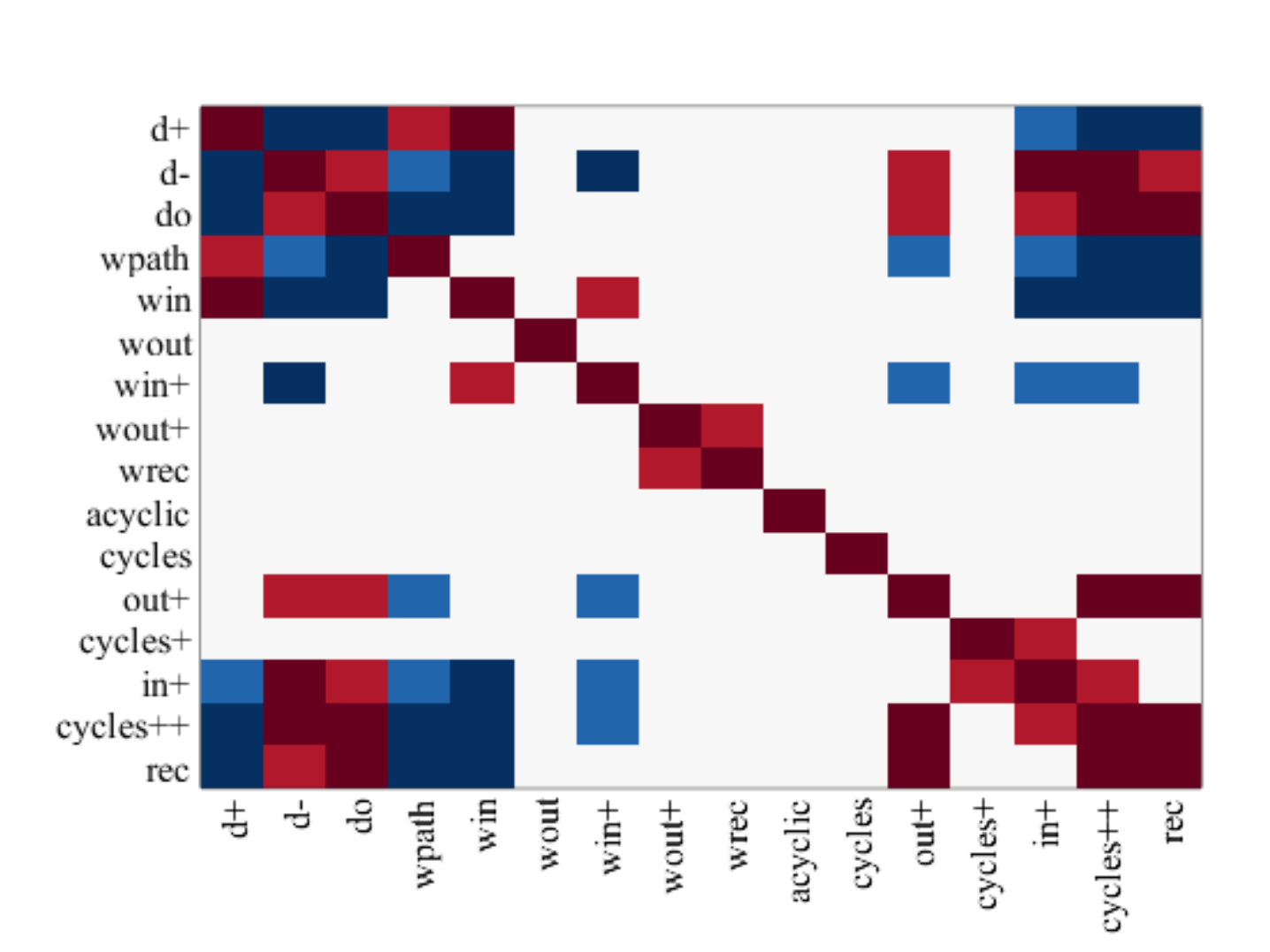}	
}%

\subfloat[Subject 27]{%
\includegraphics[trim=3cm 9cm 4cm 8cm, clip, width=\textwidth]{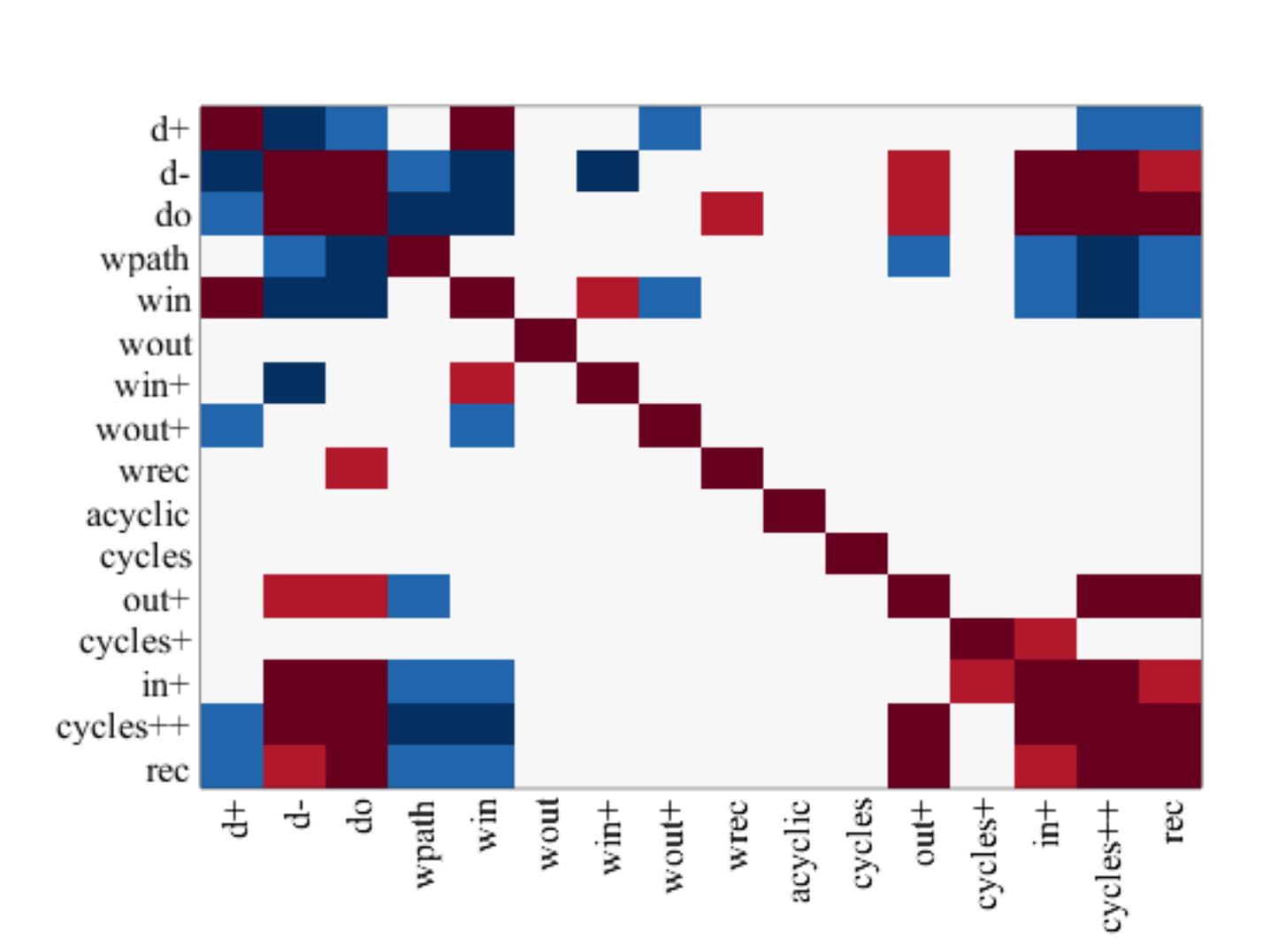}	
}%

\end{minipage}%
\begin{minipage}{.5\textwidth}
\centering
\subfloat[Subject 28]{%
\includegraphics[trim=3cm 9cm 4cm 8cm,clip,width=\textwidth]{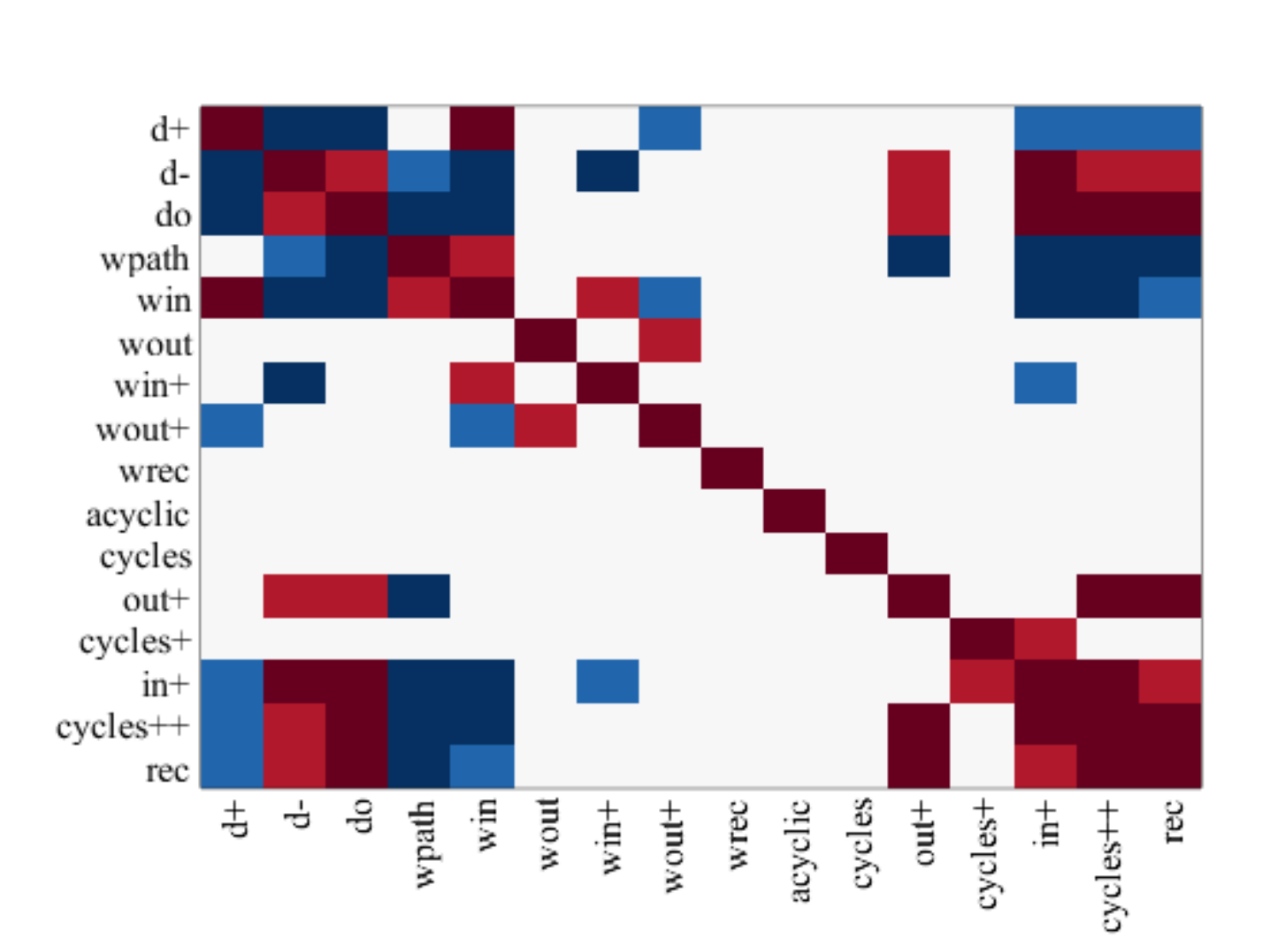}%
}%

\subfloat[Panient 29]{%
\includegraphics[trim=3cm 9cm 4cm 8cm, clip, width=\textwidth]{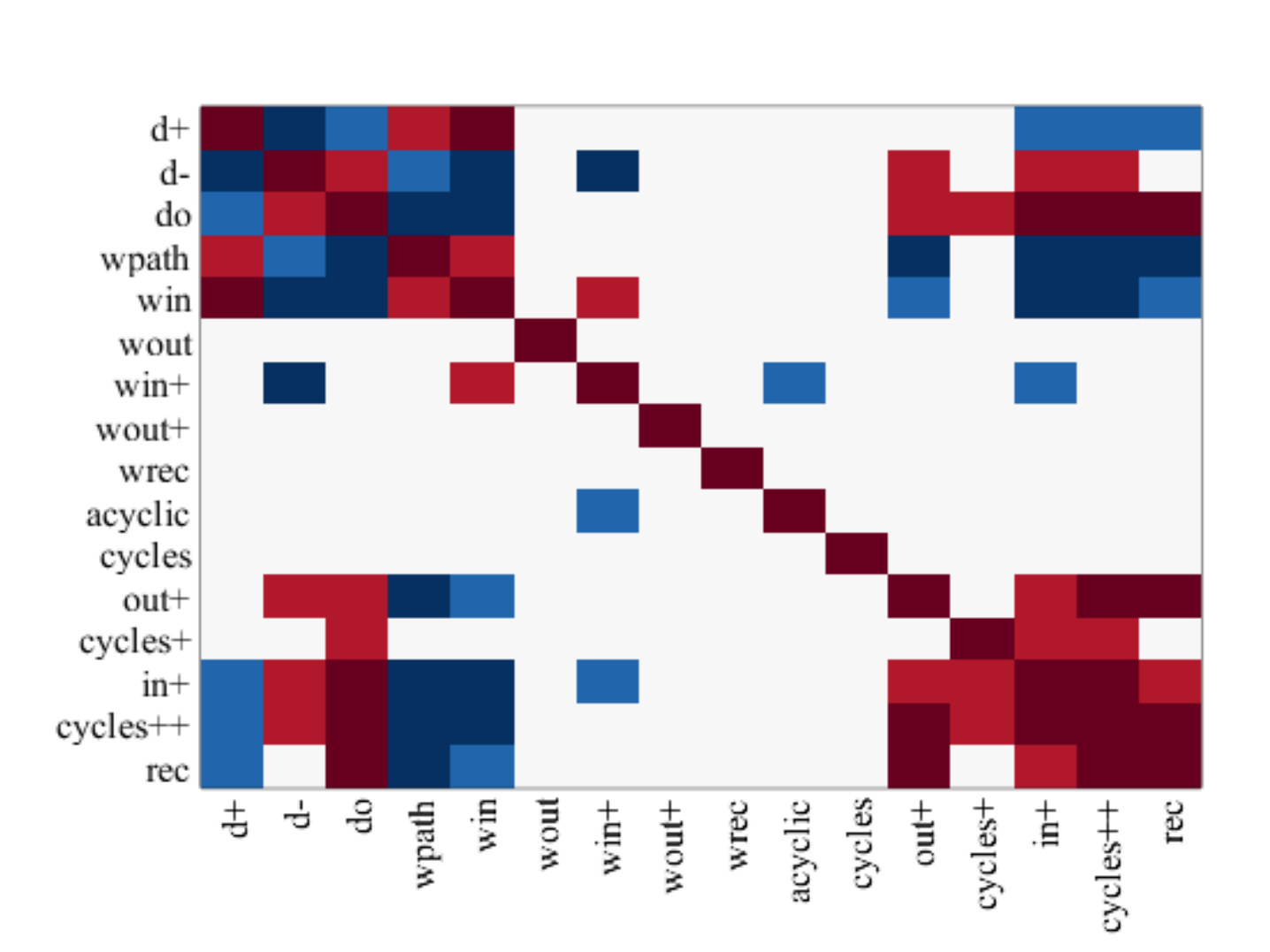}	
}%

\subfloat[Subject 30]{%
\includegraphics[trim=3cm 9cm 4cm 8cm, clip, width=\textwidth]{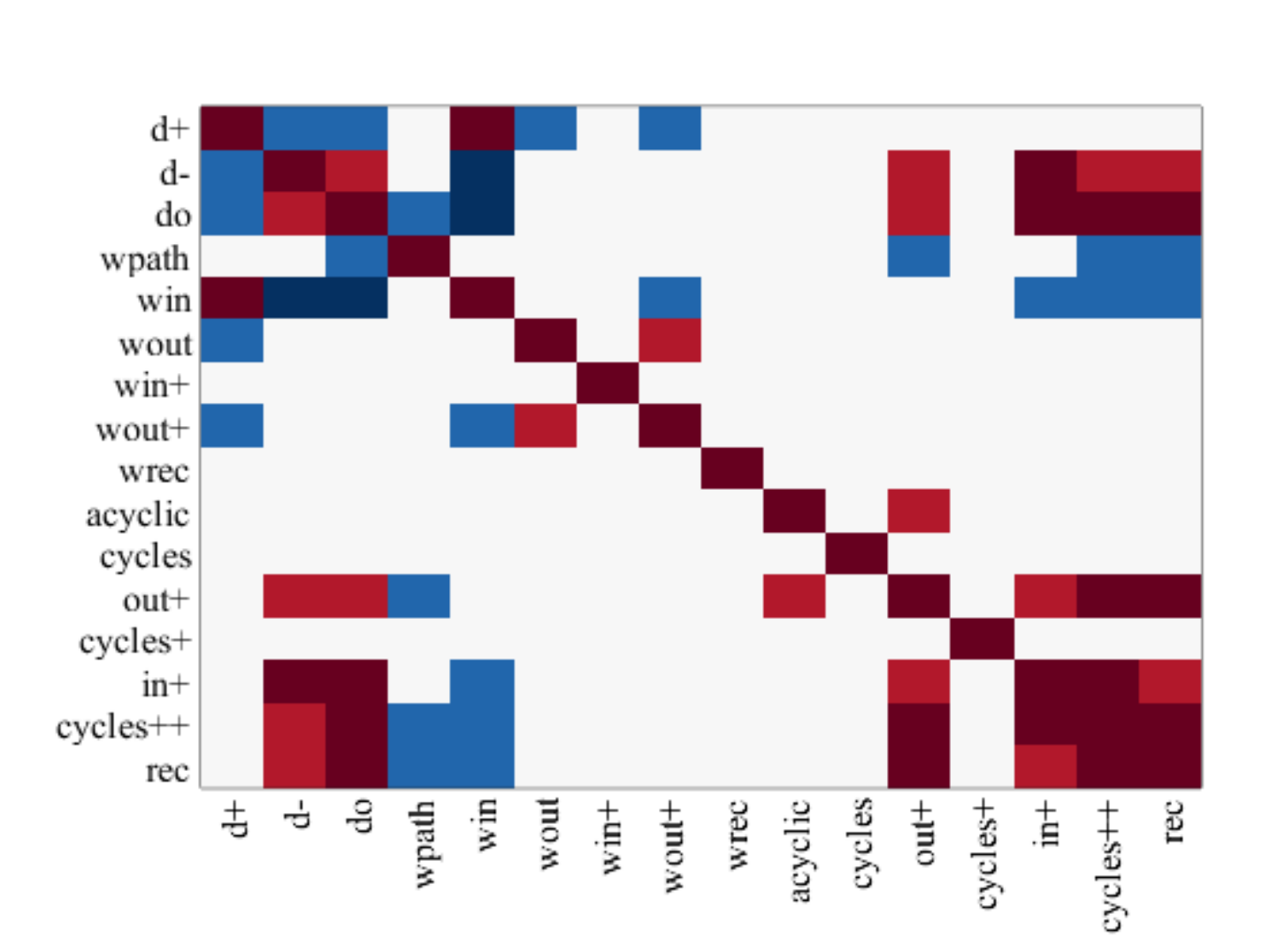}	
}%
\end{minipage}

\caption{Graphlet correlation matrices for subjects 25 to 30}
\end{figure*}

\begin{figure*}[htp]
\centering

\begin{minipage}{.5\textwidth}
\centering
\subfloat[Subject 31]{%
 \includegraphics[trim=3cm 9cm 4cm 8cm,clip,width=\textwidth]{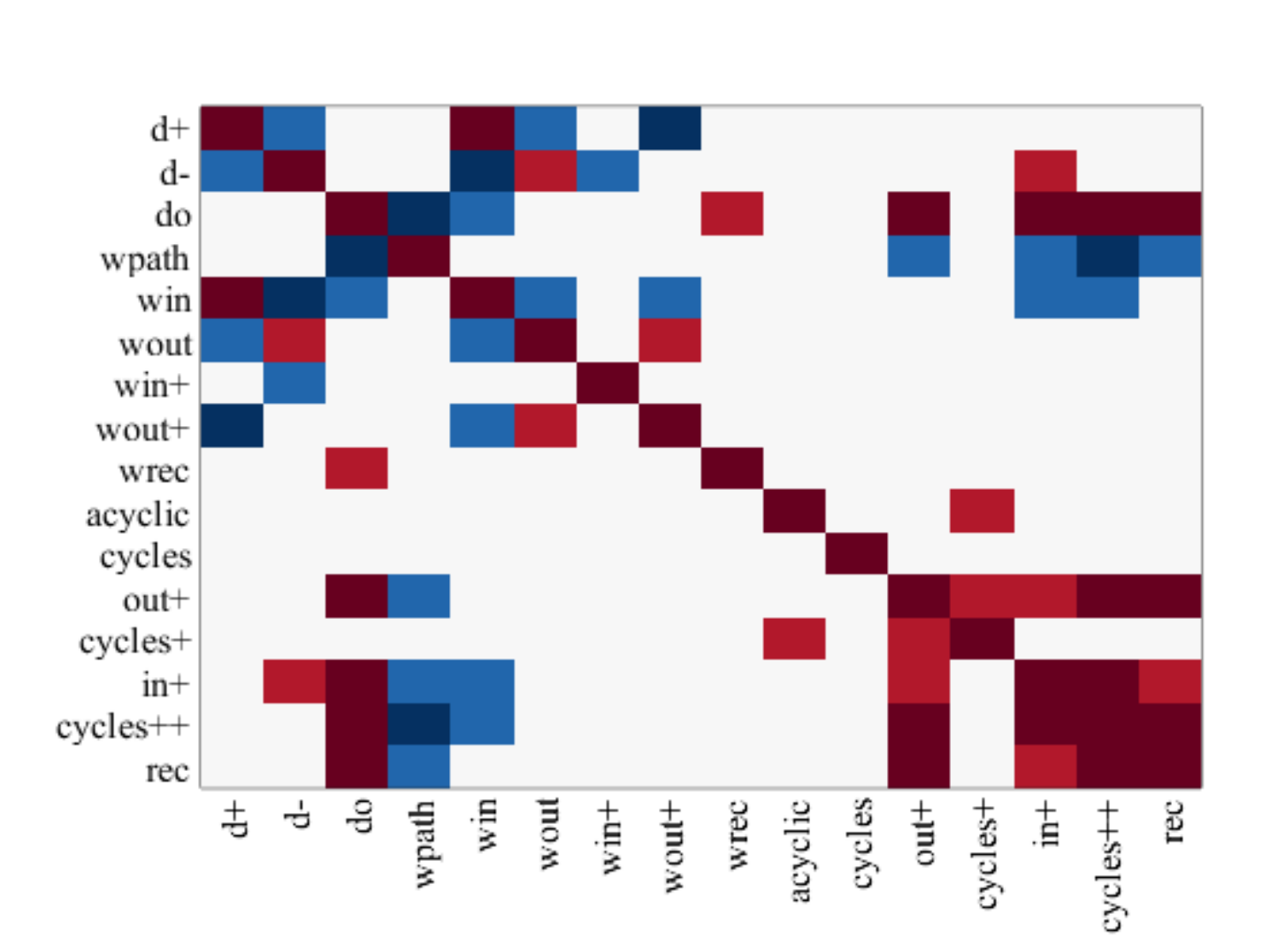}%
}

\subfloat[Subject 32]{%
\includegraphics[trim=3cm 9cm 4cm 8cm, clip, width=\textwidth]{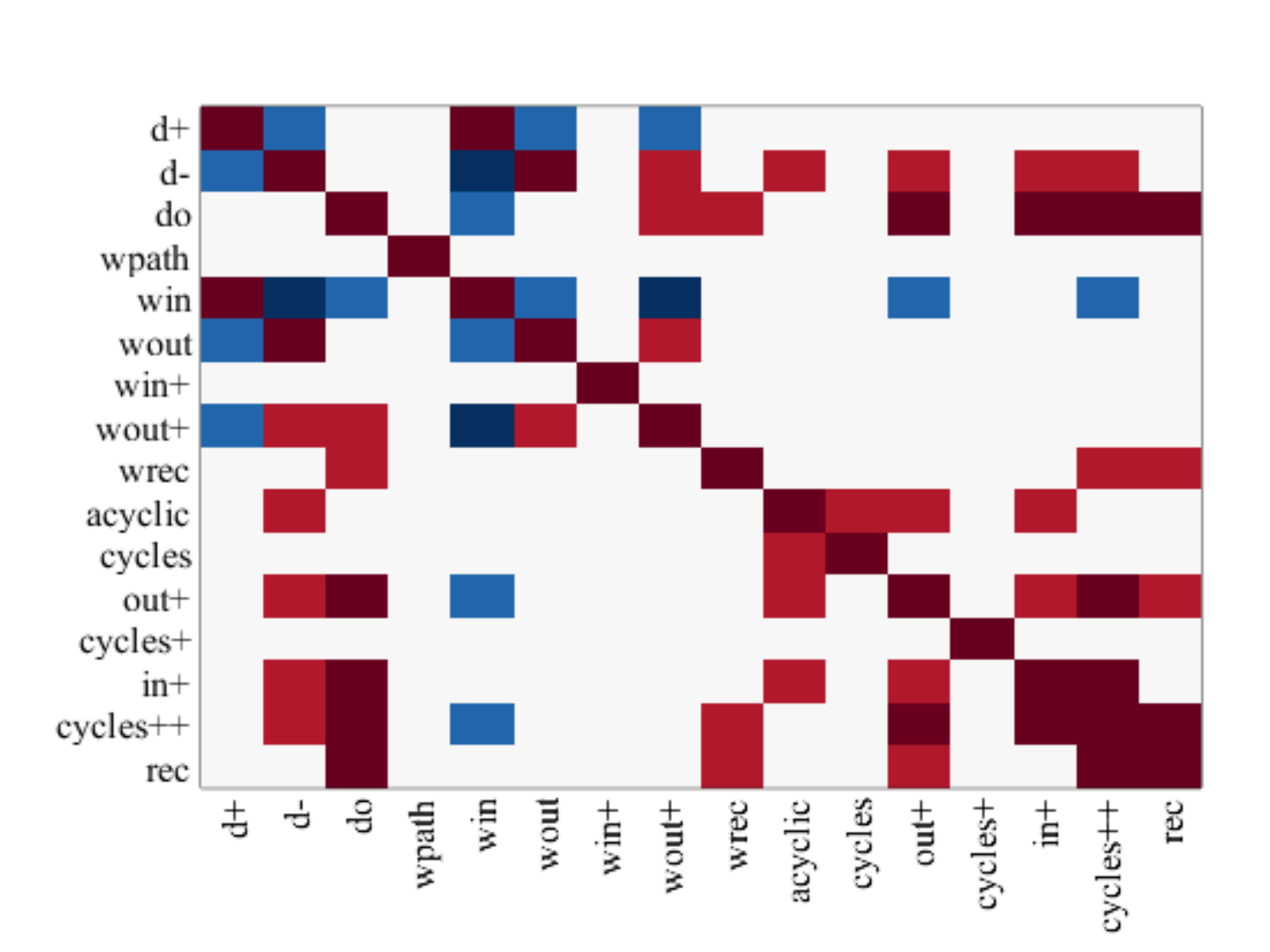}	
}%

\subfloat[Subject 33]{%
\includegraphics[trim=3cm 9cm 4cm 8cm, clip, width=\textwidth]{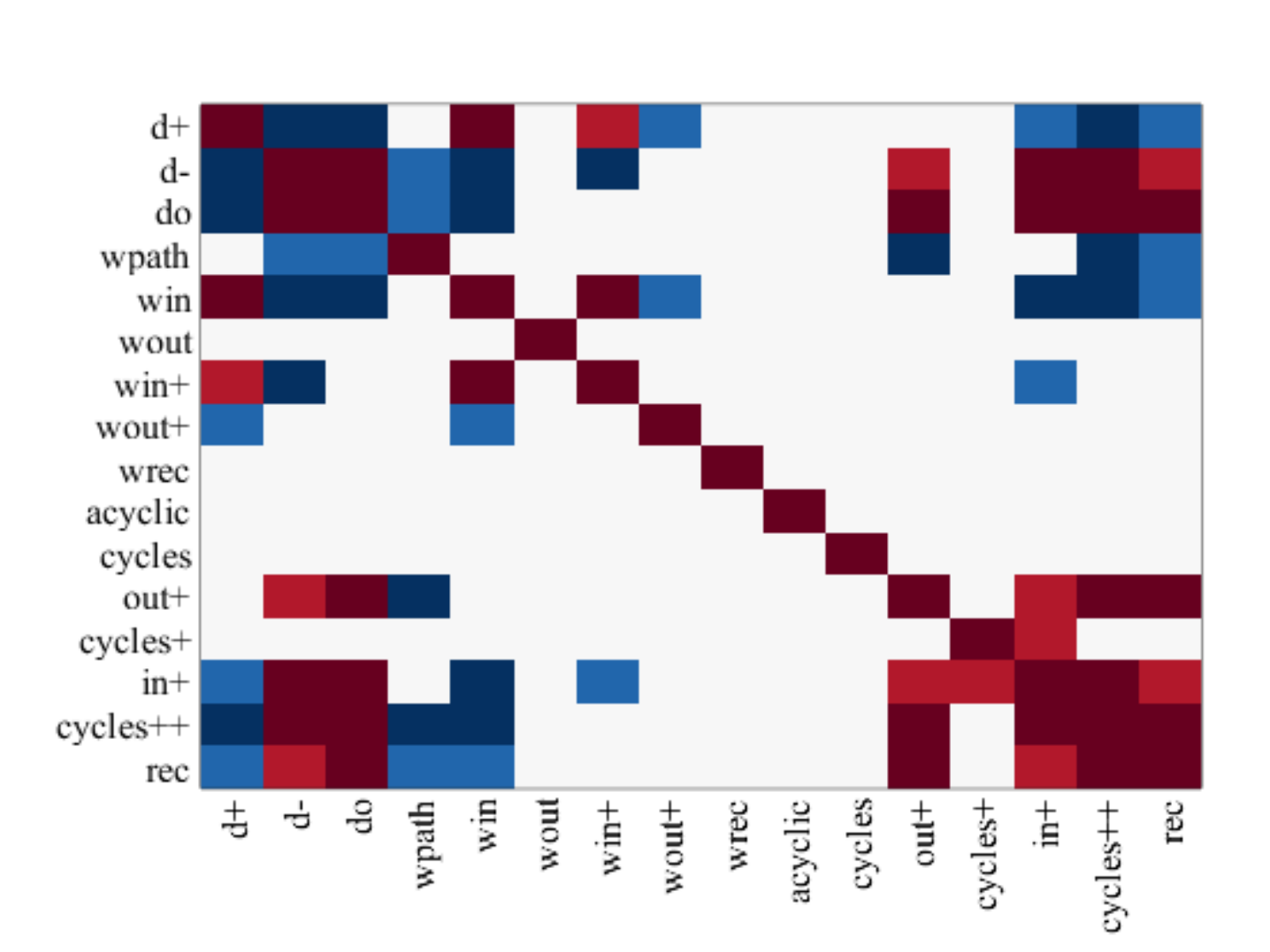}	
}%

\end{minipage}%
\begin{minipage}{.5\textwidth}
\centering
\subfloat[Subject 34]{%
\includegraphics[trim=3cm 9cm 4cm 8cm,clip,width=\textwidth]{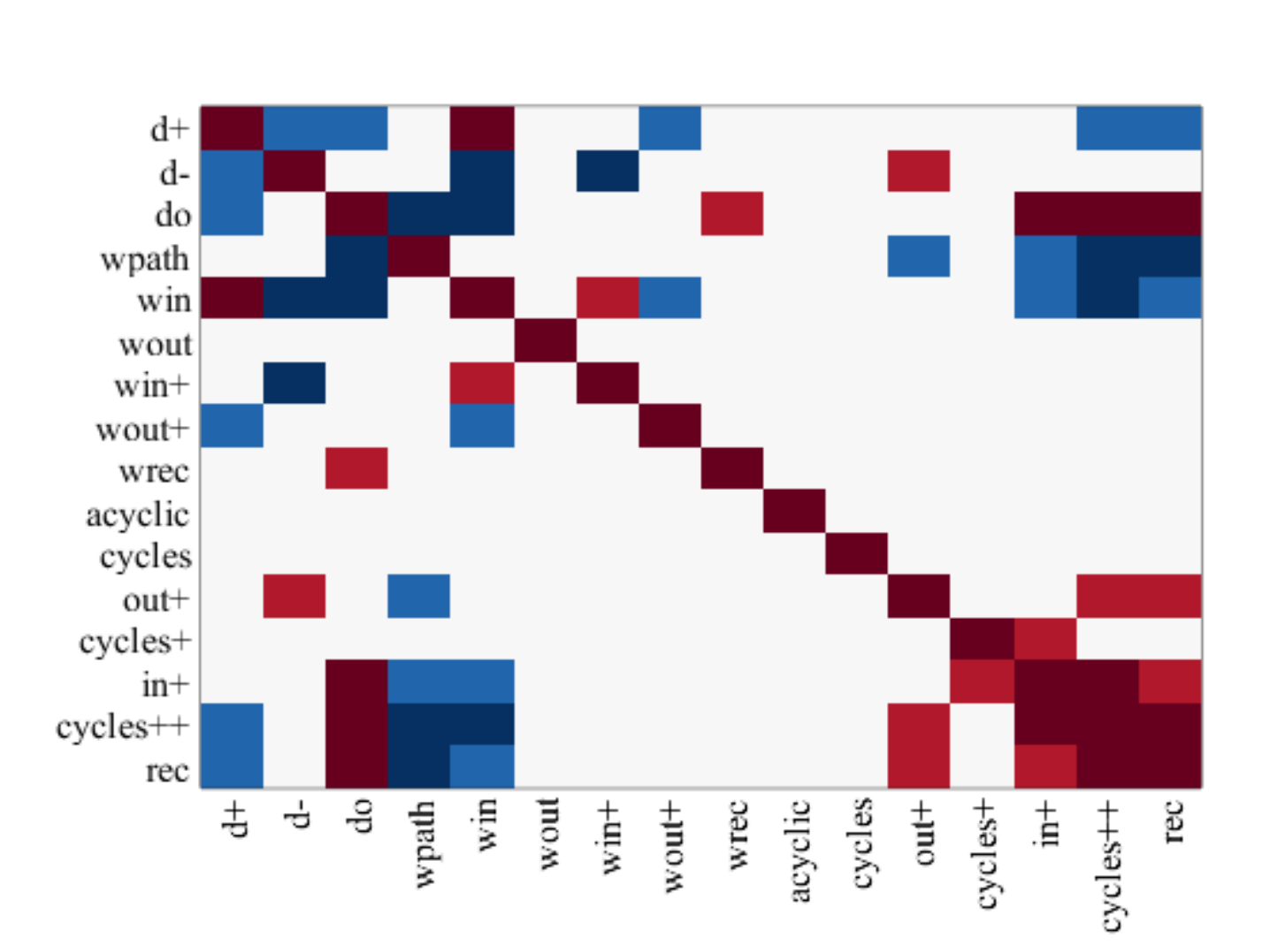}%
}%

\subfloat[Panient 35]{%
\includegraphics[trim=3cm 9cm 4cm 8cm, clip, width=\textwidth]{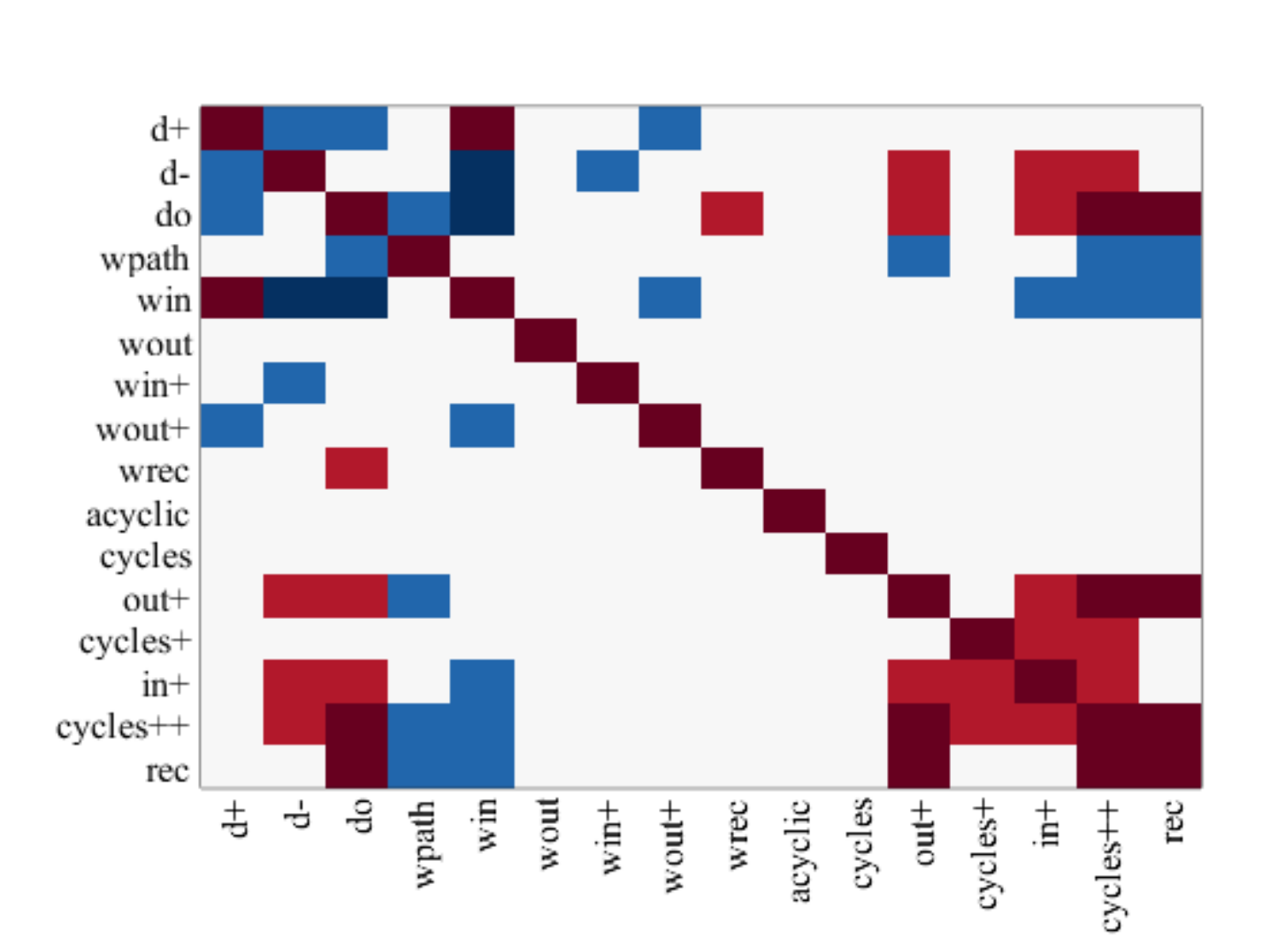}	
}%

\subfloat[Subject 36]{%
\includegraphics[trim=3cm 9cm 4cm 8cm, clip, width=\textwidth]{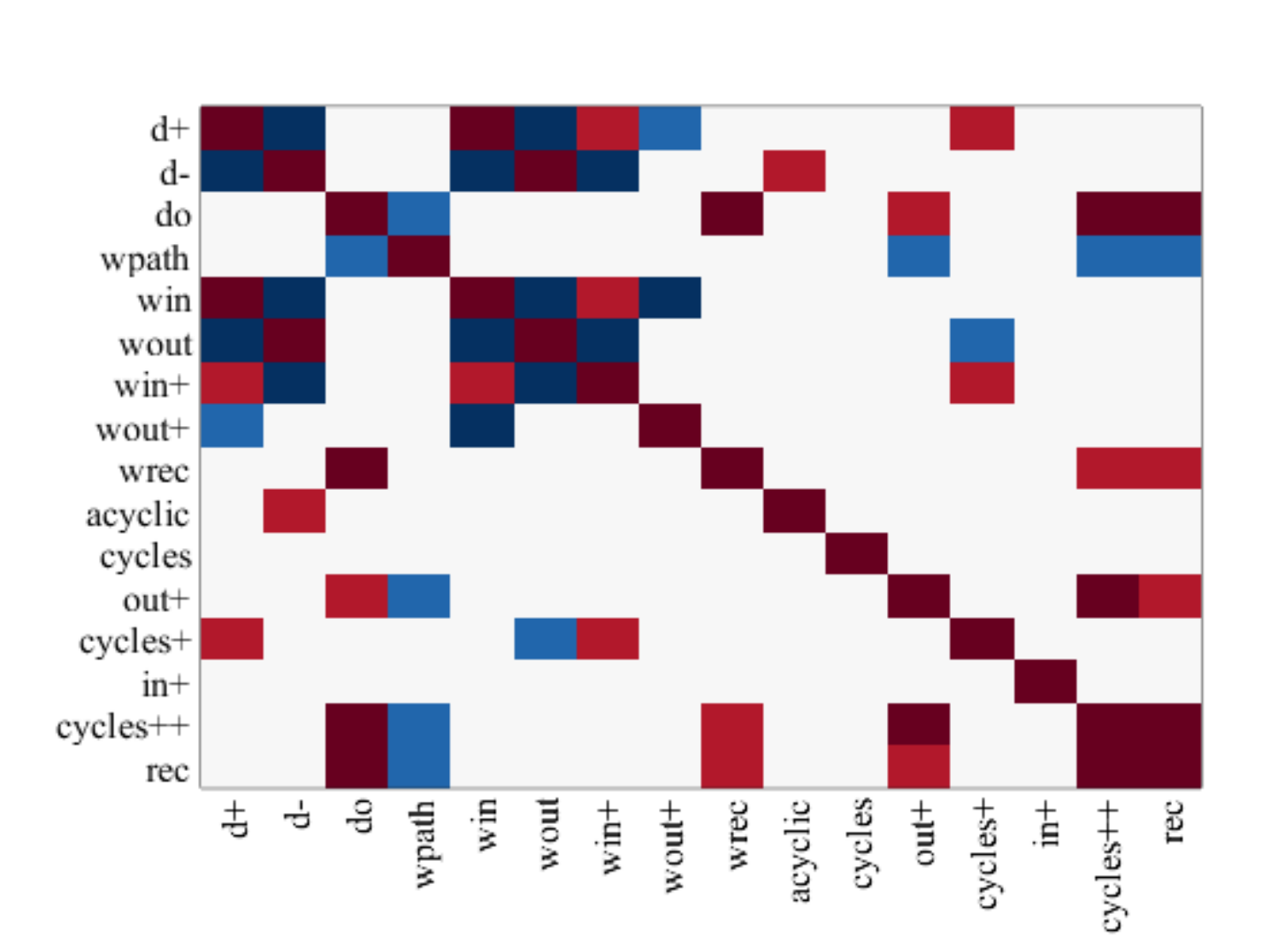}	
}%
\end{minipage}

\caption{Graphlet correlation matrices for subjects 31 to 36}
\end{figure*}

\begin{figure*}[htp]
\centering

\begin{minipage}{.5\textwidth}
\centering
\subfloat[Subject 37]{%
 \includegraphics[trim=3cm 9cm 4cm 8cm,clip,width=\textwidth]{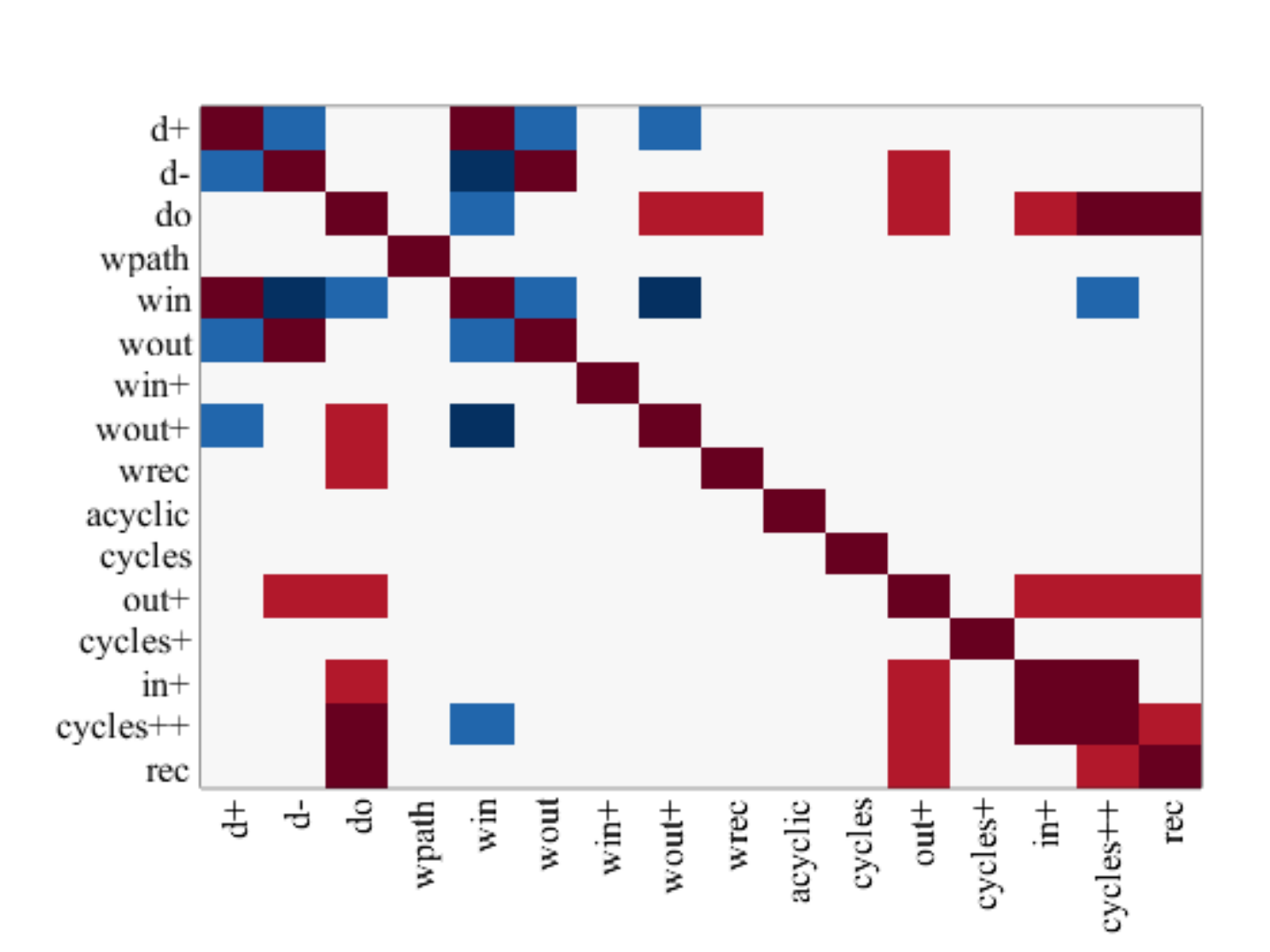}%
}

\subfloat[Subject 38]{%
\includegraphics[trim=3cm 9cm 4cm 8cm, clip, width=\textwidth]{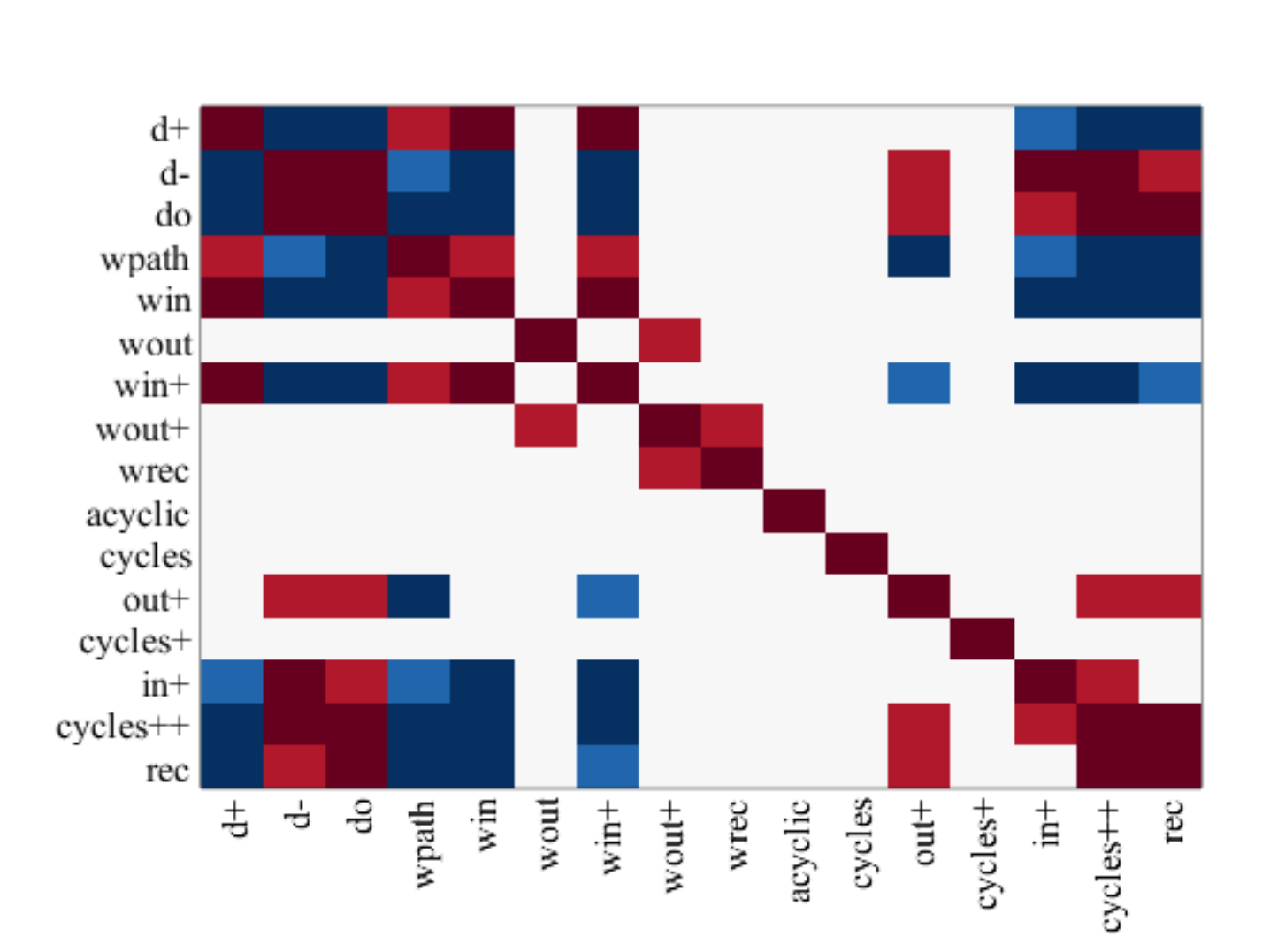}

}%

\end{minipage}%
\begin{minipage}{.5\textwidth}
\centering
\subfloat[Subject 39]{%
\includegraphics[trim=3cm 9cm 4cm 8cm,clip,width=\textwidth]{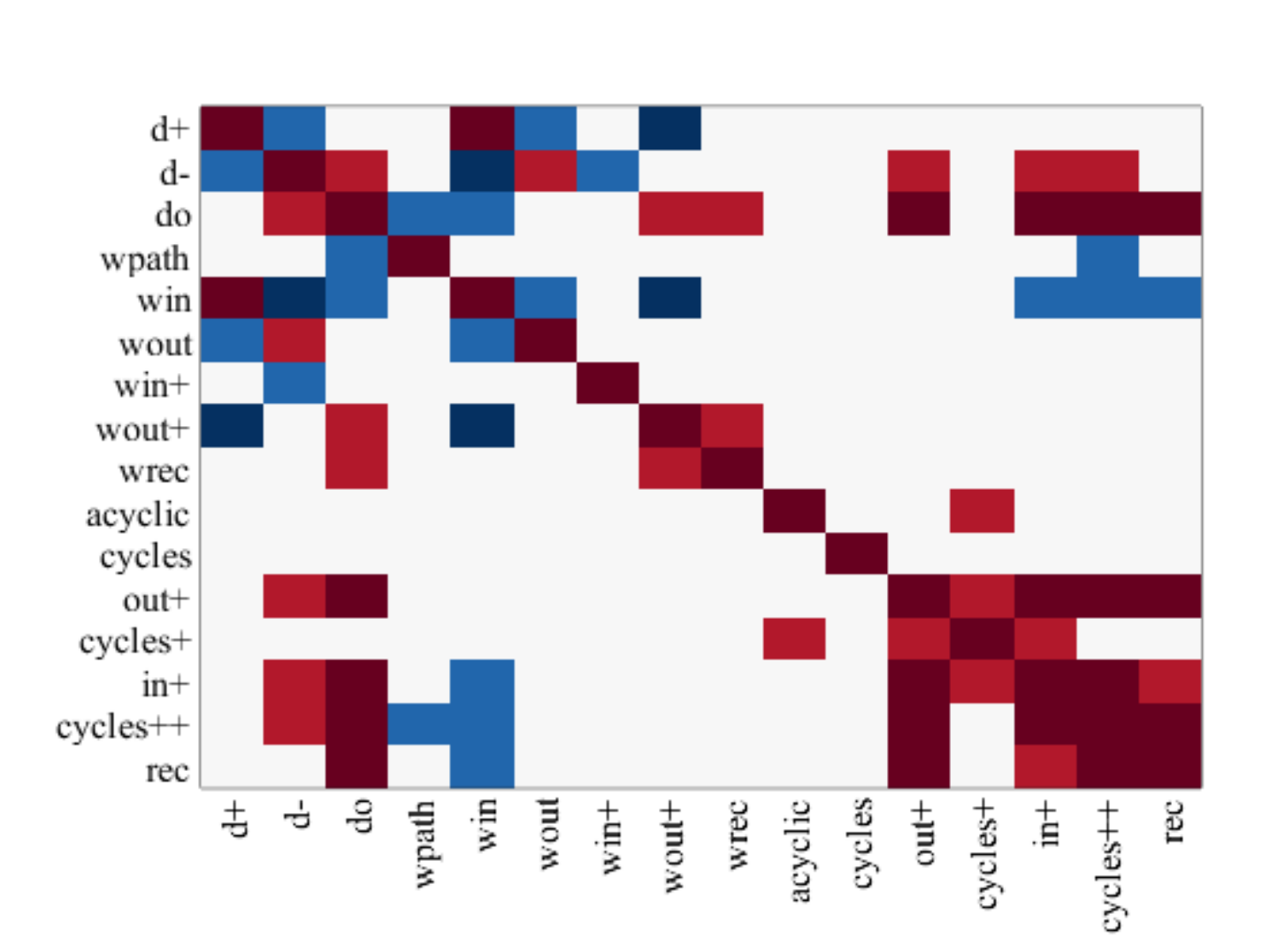}%
}%

\subfloat[Panient 40]{%
\includegraphics[trim=3cm 9cm 4cm 8cm, clip, width=\textwidth]{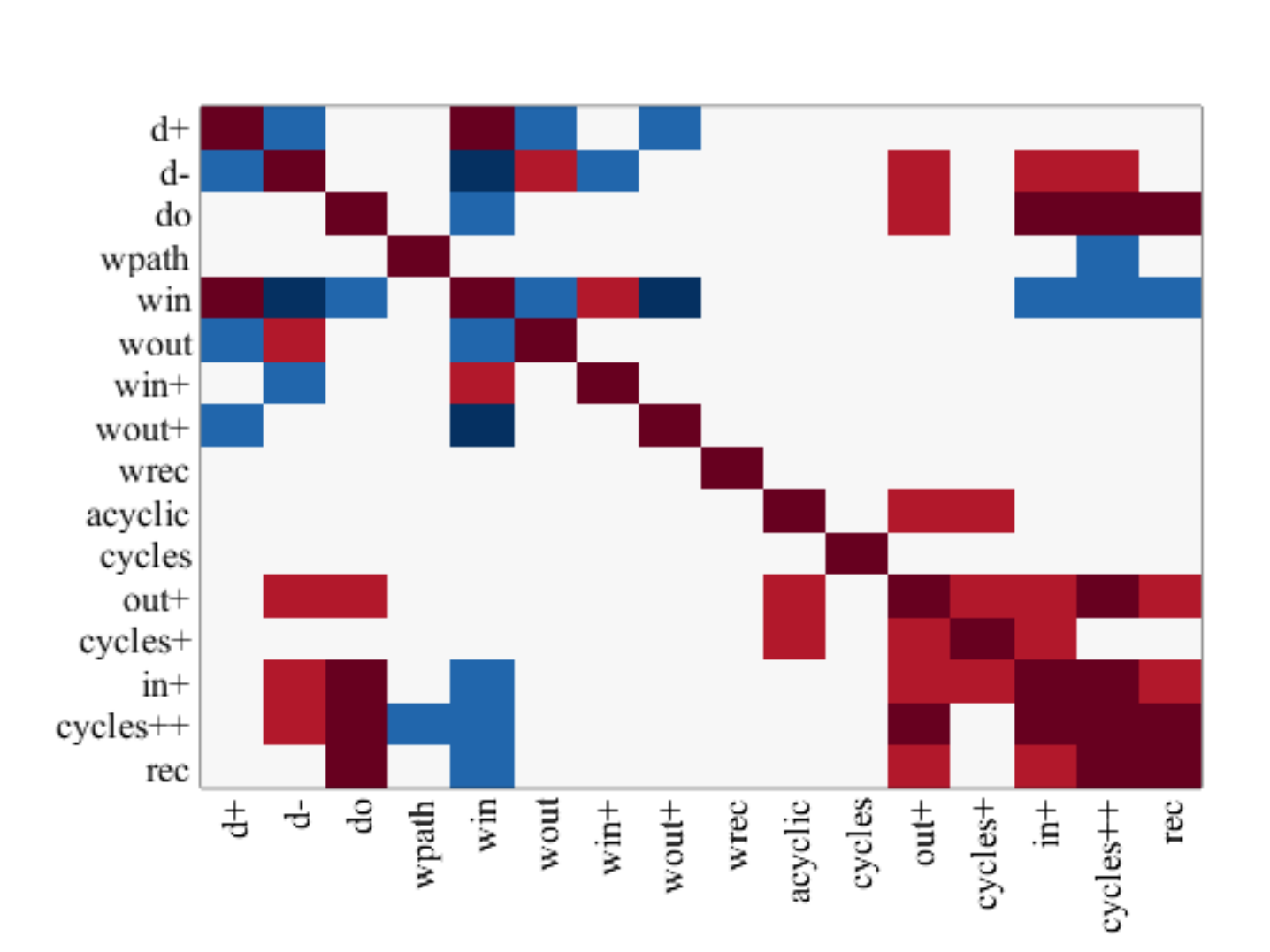}	
}%
\end{minipage}

\caption{Graphlet correlation matrices for subjects 37 to 40}
\end{figure*}

\end{document}